\documentclass[english,aps,prd,amsmath,amssymb,amsfonts,10pt,onecolumn,a4paper,notitlepage]{revtex4}
\usepackage[T1]{fontenc}
\usepackage[latin1]{inputenc}
\usepackage{bm}
\usepackage{amsmath}
\usepackage{amssymb}
\usepackage{graphicx}
\usepackage{esint}

\makeatletter

\@ifundefined{textcolor}{}
{%
 \definecolor{BLACK}{gray}{0}
 \definecolor{WHITE}{gray}{1}
 \definecolor{RED}{rgb}{1,0,0}
 \definecolor{GREEN}{rgb}{0,1,0}
 \definecolor{BLUE}{rgb}{0,0,1}
 \definecolor{CYAN}{cmyk}{1,0,0,0}
 \definecolor{MAGENTA}{cmyk}{0,1,0,0}
 \definecolor{YELLOW}{cmyk}{0,0,1,0}
 }

\usepackage{dcolumn}\usepackage{bm}

\def\be{\begin{equation}}
\def\ee{\end{equation}}
\def\ba{\begin{eqnarray}}
\def\ea{\end{eqnarray}}
\def\bs{\begin{subequations}}
\def\es{\end{subequations}}

\def\tx{\tilde{x}}
\def\ty{\tilde{y}}

\@ifundefined{definecolor}
 {\usepackage{color}}{}

\newcommand{\GB}{\mathcal{G}}
\newcommand{\de}{\mathrm{d}}

\makeatother

\begin{document}

\title{Generalized Galileon cosmology}

\author{Antonio \surname{De Felice}}

\affiliation{Department of Physics, Faculty of Science, Tokyo University of Science,
1-3, Kagurazaka, Shinjuku-ku, Tokyo 162-8601, Japan}

\author{Shinji \surname{Tsujikawa}}

\affiliation{Department of Physics, Faculty of Science, Tokyo University of Science,
1-3, Kagurazaka, Shinjuku-ku, Tokyo 162-8601, Japan}

\begin{abstract}

We study the cosmology of a generalized Galileon field $\phi$ with
five covariant Lagrangians in which $\phi$ is replaced by general
scalar functions $f_{i}(\phi)$ ($i=1,\cdots,5$). For these theories,
the equations of motion remain at second-order in time derivatives.
We restrict the functional forms of $f_{i}(\phi)$ from the demand
to obtain de Sitter solutions responsible for dark energy. There are
two possible choices for power-law functions $f_{i}(\phi)$, depending
on whether the coupling $F(\phi)$ with the Ricci scalar $R$ is independent
of $\phi$ or depends on $\phi$. The former corresponds to the covariant
Galileon theory that respects the Galilean symmetry in the Minkowski
space-time. For generalized Galileon theories we derive the conditions
for the avoidance of ghosts and Laplacian instabilities associated
with scalar and tensor perturbations as well as the condition for
the stability of de Sitter solutions. We also carry out detailed analytic
and numerical study for the cosmological dynamics in those theories.

\end{abstract}

\date{\today}

\maketitle

%%%%%%%%%%%%%%
\section{Introduction}
%%%%%%%%%%%%%%

The $\Lambda$-Cold-Dark-Matter ($\Lambda$CDM) model has been consistent
with observational data, but the energy scale of dark energy is too
low to be compatible with the cosmological constant originated from
the vacuum energy in quantum field theory \cite{Weinberg}. Since
the observations allow the variation of the dark energy equation of
state \cite{obserpapers}, many models have been proposed to explain
the present accelerated expansion of the Universe \cite{darkreview}.
For example, a light scalar field with a slowly varying potential,
called quintessence, was introduced as an alternative to the cosmological
constant \cite{quin}. In general, however, it is not easy to construct
viable particle physics models of quintessence because of an extremely
light mass required for the cosmic acceleration today \cite{Carroll}.

Another approach for addressing the dark energy problem is to modify
the law of gravity from General Relativity at large distances \cite{modifiedreview}.
In this approach there have been two main streams. The first consists
of introducing a Lagrangian for gravity built up out of the Ricci,
Riemann, and metric tensors, which generally leads to 4-th order differential
equations. The $f(R)$ gravity \cite{fR} and the Gauss-Bonnet gravity
\cite{fG} belong to this class. The second consists of higher dimensional
models that realize the cosmic acceleration through the gravitational
leakage to extra dimensions. The Dvali-Gabadadze-Porrati (DGP) braneworld
model \cite{DGP} belongs to this class (see Refs.~\cite{DGPge}).

In general, modified gravity models of dark energy need to be constructed
to recover the General Relativistic behavior in the regions of high
density for the consistency with local gravity experiments. In $f(R)$
gravity, there have been a number of viable models in which a scalar-field
degree of freedom ({}``scalaron'' \cite{scalaron}) has a large
mass in the region where the Ricci scalar $R$ is much larger than
its cosmological value $R_{0}$ today \cite{fR2}. Provided that the
chameleon mechanism \cite{chameleon} is at work in the local regime,
the gravitational coupling with non-relativistic matter can be suppressed
to be compatible with solar system experiments \cite{localfR}. There
is also another mechanism called the Vainshtein screen effect \cite{Vainshtein}
in which non-linear effects can effectively decouple the scalar field
from gravity. Originally the Vainshtein mechanism was applied to the
theories of massive gravity like Fierz-Pauli gravity \cite{Pauli}
(see also Refs.~\cite{Babichev}), but the non-linearities imply
the presence of a ghost state in such theories \cite{Cremi}.

In the DGP model non-linear field self-interacting Lagrangians such
as $\square\phi(\partial_{\mu}\phi\partial^{\mu}\phi)$ arise from
a brane-bending mode (i.e. a longitudinal graviton) \cite{DGPnon}.
This allows the decoupling of $\phi$ from gravitational dynamics
in the local region. Unfortunately the self-accelerating solution
in the DGP model contains a ghost mode \cite{DGPghost} even in the
absence of non-linear terms. Moreover the model is disfavored from
the combined data analysis of supernovae Ia and baryon acoustic oscillations
\cite{DGPobser}.

Mostly inspired by the DGP model, Nicolis \textit{et al.} \cite{Nicolis}
derived the five Lagrangians that lead to the field equations invariant
under the Galilean symmetry $\partial_{\mu}\phi\to\partial_{\mu}\phi+b_{\mu}$
in the Minkowski space-time {[}including the term $\square\phi(\partial_{\mu}\phi\partial^{\mu}\phi)${]}.
The scalar field that respects the Galilean symmetry is dubbed {}``Galileon''.
Each of the five terms only leads to second-order differential equations,
keeping the theory free from unstable spin-2 ghost degrees of freedom.
If we extend the analysis in Ref.~\cite{Nicolis} to the curved space-time,
the Lagrangians need to be promoted to the covariant forms. Deffayet
\textit{et al.} \cite{DeffaGal,DeffaGalED} derived the covariant
Lagrangians ${\cal L}_{i}$ ($i=1,\cdots,5$) that keep the field
equations up to second-order. We caution that in the curved space-time
the Galilean symmetry is in general broken for non-linear field self-interacting
terms, but in the Minkowski space-time it is preserved for the covariant
Lagrangians ${\cal L}_{i}$ ($i=1,\cdots,5$).

The (modified) Galileon gravity has been extensively applied to cosmology recently \cite{JustinGal,KazuyaGal,Kobayashi1,Kobayashi2,Rham10,Sami,DT1,DMT,Cremi2,DT2,Padilla,Deser,DPSV,KYY,Mark,Ali,Nesseris}.
One application is to introduce the non-linear field self-interaction
of the form $\xi(\phi)\square\phi(\partial_{\mu}\phi\partial^{\mu}\phi)$
in the action of (generalized) Brans-Dicke theories \cite{KazuyaGal,Kobayashi1,Kobayashi2,DT1,DMT},
where $\xi$ is a function of $\phi$. Although such a term breaks
the Galilean symmetry, the field equations remain at second-order.
Moreover, for suitable choices of the function $\xi(\phi)$, there
exist de Sitter (dS) solutions responsible for dark energy even in
the absence of the field potential. The presence of the non-linear
term also allows the decoupling of the field from gravity in the regions
of high density under the Vainshtein mechanism.

Another application of Galileon gravity to cosmology is to study the
expansion history of the Universe in the presence of the covariant
Lagrangians ${\cal L}_{i}$ ($i=1,\cdots,5$) mentioned above. The
cosmology up to the term ${\cal L}_{4}$ has been discussed in Ref.~\cite{Sami},
which showed the existence of stable dS solutions. Recently the full
cosmological dynamics including the term ${\cal L}_{5}$ have been
studied in Ref.~\cite{DT2}. The viable model parameter space has
been found by studying the conditions for the avoidance of ghosts
and Laplacian instabilities. Interestingly there exists a tracker
solution that finally approaches a stable dS solution. The equation
of state of dark energy exhibits a peculiar phantom-like behavior
along the tracker.

In this paper we shall study general Galileon theories in which the
field $\phi$ in the covariant Lagrangians ${\cal L}_{i}$ ($i=1,\cdots,5$)
is replaced by general functions $f_{i}(\phi)$. Since $f_{i}(\phi)$
are scalar functions, the resulting field equations also remain at
second-order. We constrain the forms of $f_{i}(\phi)$ from the requirement
to obtain dS solutions. This constraint gives rise to the Galileon
theory with $f_{i}(\phi)\propto\phi$ as a specific case. For general
functions $f_{i}(\phi)$ we also derive the conditions for the avoidance
of ghosts and Laplacian instabilities. This is useful to constrain
the viable parameter space of those theories. 
We shall perform detailed analytic and numerical study for the
cosmological dynamics of generalized Galileon theory 
with several different choices of $f_{i}(\phi)$.

%%%%%%%%%%%%%%%%%%%%%%%%%%%%%%%%%%%%%%%%
\section{Generalized Galileon theories}
%%%%%%%%%%%%%%%%%%%%%%%%%%%%%%%%%%%%%%%%

In the curved space-time the Galilean symmetry is broken even for
the Lagrangian ${\cal L}_{2}=(\nabla\phi)^{2}\equiv
\partial_{\mu}\phi\partial^{\mu}\phi$.
Then this symmetry is not restrictive when we study the covariant
generalization of the Galileon field. On the other hand, the covariant
Galileon formalism leads to second-order field equations. We
study a Lagrangian that gives second-order equations of motion, 
such that the theories recover the covariant Lagrangian  
in Refs.~\cite{DeffaGal,DeffaGalED} as a specific case.
We will consider two generalizations of the covariant
Galileon theory: (i) scalar couplings with both the Ricci scalar $R$
and the Gauss-Bonnet (GB) term ${\cal G}$ are introduced, (ii) the
covariant Galileon terms are extended to more general functions.

As for the first point, this step is compatible with the approach
of field theory, because such scalar couplings generally exist and
they are consistent with general covariance (and even with the Galileon
symmetry, as in the Minkowski background their contributions to the
equations of motion of the field identically vanish). Moreover the
scalar couplings with $R$ and ${\cal G}$ give only second-order
contributions. It is true that the GB term can change the Ultra-Violet
behavior for the modes, but this property also holds for all the remaining
terms coming from the extended Galileon action.

As for the second point, we can replace the scalar field $\phi$ in
each Lagrangian term with a function of the field itself. The Lagrangian
${\cal L}_{2}$, for example, can be modified to 
$(\nabla f_{2}(\phi))^{2}\equiv\partial_{\mu}f_{2}(\phi)\,\partial^{\mu}f_{2}(\phi)$.
The equations still remain at second-order because $f_{2}$ is a scalar
quantity itself. We will consider this generalization for all the
Galileon terms, introducing different functions $f_{i}(\phi)$ ($i=1,2,\cdots$)
for each of them.

According to the above prescription, we introduce the following Lagrangians
as the generalization of those introduced by Deffayet \textit{et al.}
\cite{DeffaGal}: 
\begin{eqnarray}
 &  & \hspace{-0.5cm}{\cal L}_{1}=f_{1}(\phi)\,,\\
 &  & \hspace{-0.5cm}{\cal L}_{2}=(\nabla f_{2}(\phi))^{2}\,,\\
 &  & \hspace{-0.5cm}{\cal L}_{3}=(\Box f_{3}(\phi))\,(\nabla f_{3}(\phi))^{2}\,,\\
 &  & \hspace{-0.5cm}{\cal L}_{4}=(\nabla f_{4}(\phi))^{2}\left[2(\Box f_{4}(\phi))^{2}-2f_{4}(\phi)_{;\mu\nu}f_{4}(\phi)^{;\mu\nu}-(R/2)\,(\nabla f_{4}(\phi))^{2}\right]\,,\\
 &  & \hspace{-0.5cm}{\cal L}_{5}=(\nabla f_{5}(\phi))^{2}\,[(\Box f_{5}(\phi))^{3}-3(\Box f_{5}(\phi))f_{5}(\phi)_{;\mu\nu}f_{5}(\phi)^{;\mu\nu}+2\, f_{5}(\phi)_{;\mu}{}^{\nu}f_{5}
 (\phi)_{;\nu}{}^{\rho}f_{5}(\phi)_{;\rho}{}^{\mu}\nonumber \\
 &  & ~~~~~~~~~~~~~~~~~~-6\, f_{5}(\phi)_{;\mu}f_{5}(\phi)^{;\mu\nu}f_{5}(\phi)^{;\rho}G_{\nu\rho}],
\end{eqnarray}
 where $R$ is the Ricci scalar and $G_{\nu\rho}$ is the Einstein
tensor. One can also introduce the following terms 
\begin{eqnarray}
{\cal L}_{6} & = & F(\phi)\, R\,,\label{eq:FnmR}\\
{\cal L}_{7} & = & \xi(\phi)\,\GB\,,\label{eq:XinmG}
\end{eqnarray}
 which vanish in the Minkowski space-time. Here $\GB=R^{2}-4R_{\alpha\beta}R^{\alpha\beta}+R_{\alpha\beta\gamma\delta}R^{\alpha\beta\gamma\delta}$
represents the GB combination, where $R_{\alpha\beta}$ is the Ricci
tensor and $R_{\alpha\beta\gamma\delta}$ is the Riemann tensor.

The covariant action we shall discuss is given by 
\begin{equation}
S=\frac{1}{2}\int{\rm d}^{4}x\sqrt{-g}\,{\cal L}+\int{\rm d}^{4}x\,{\cal L}_{M}\,,\label{action}
\end{equation}
 where $g$ is a determinant of the space-time metric $g_{\mu\nu}$,
${\cal L}_{M}$ is a matter Lagrangian, and 
\begin{equation}
{\cal L}=F(\phi)R+\xi(\phi)\GB+{\cal L}_{1}+\epsilon_{2}{\cal L}_{2}+{\cal L}_{3}+\epsilon_{4}{\cal L}_{4}+{\cal L}_{5}\,.\label{Lag}
\end{equation}
 In order to control the signs of ${\cal L}_{2,4}$ we have introduced
the factors $\epsilon_{2,4}$, which are either $+1$ or $-1$. For
the terms ${\cal L}_{1,3,5}$ we get the opposite signs by replacing
$f_{i}(\phi)$ with $-f_{i}(\phi)$. For the matter Lagrangian ${\cal L}_{M}$
we consider the contribution of two perfect fluids ${\cal L}_{M}^{(i)}$
($i=1,2$), described by the barotropic equations of state of the
form $w_{i}=P_{i}/\rho_{i}$ ($i=1,2$). Note that $P_{i}$ and $\rho_{i}$
are the pressure and the energy density of fluids, with the energy-momentum
tensor $T_{\mu\nu}^{(i)}=-(2/\sqrt{-g})\delta{\cal L}_{M}^{(i)}/\delta g^{\mu\nu}$.

%%%%%%%%%%%%%%%%%%%%%%%%%%%%%%
\section{Background cosmology}
\label{cosmosec} 
%%%%%%%%%%%%%%%%%%%%%%%%%%%%%%

Consider the flat Friedmann-Lemaitre-Robertson-Walker (FLRW) 
space-time with the line element 
\begin{equation}
\de s^{2}=g_{\mu\nu}{\rm d}x^{\mu}{\rm d}x^{\nu}=-\de t^{2}+a^{2}(t)\,\de\bm{x}^{2}\,,\label{eq:flrw}
\end{equation}
 where $a(t)$ is the scale factor with cosmic time $t$. Varying
the action (\ref{action}) with respect to $g_{\mu\nu}$, we obtain
the following equations of motion 
\begin{eqnarray}
&  & 3FH^{2}+3H\dot{F}+f_{1}/2+\epsilon_{2}\dot{f}_{2}^{2}/2
-3H\dot{f}_{3}^{3}+45\epsilon_{4}H^{2}\dot{f}_{4}^{4}/2-21H^{3}
\dot{f}_{5}^{5}+12H^{3}\dot{\xi}=\rho_{1}+\rho_{2}\,,
\label{eq:frd1}\\
&  & 3FH^{2}+\ddot{F}+2H\dot{F}+2F\dot{H}+f_{1}/2-\epsilon_{2}
\dot{f}_{2}^{2}/2-\dot{f}_{3}^{2}\ddot{f}_{3}+3\epsilon_{4}
\dot{f}_{4}^{3}(8H\ddot{f}_{4}+3H^{2}\dot{f}_{4}+2\dot{H}
\dot{f}_{4})/2\nonumber \\
&  & -3H\dot{f}_{5}^{4}(5H\ddot{f}_{5}+2\dot{H}\dot{f}_{5}
+2H^{2}\dot{f}_{5})+4H[H\ddot{\xi}+2(H^{2}+\dot{H})\dot{\xi}]
=-(w_{1}\rho_{1}+w_{2}\rho_{2})\,,
\label{eq:frd2}
\end{eqnarray}
 where a dot represents a derivative with respect to $t$. The matter
fluids obey the continuity equations 
\begin{equation}
\dot{\rho}_{i}+3H(1+w_{i})\rho_{i}=0\,,\qquad(i=1,2).\label{conser}
\end{equation}
 Differentiating Eq.~(\ref{eq:frd1}) in terms of $t$ and eliminating
the terms $w_{1}\rho_{1}+w_{2}\rho_{2}$ from Eq.~(\ref{eq:frd2}),
we get the generalized Klein-Gordon equation for the scalar field.
For the perfect fluids we consider radiation ($w_{1}=1/3$, $\rho_{1}=\rho_{r}$)
and non-relativistic matter ($w_{2}=0$, $\rho_{2}=\rho_{m}$).

Let us restrict the functional forms of $f_{i}(\phi)$, $F(\phi)$,
and $\xi(\phi)$ by demanding the existence of dS solutions responsible
for dark energy. We shall focus on the theories in which the late-time
cosmic acceleration can be realized by the field kinetic terms rather
than the field potential, so that we set 
\begin{equation}
f_{1}(\phi)=0\,.\label{eq:nopotential}
\end{equation}
 The absence of this term and more in general of a potential for the
field can be implemented by invoking an additional shift symmetry
$\phi\to\phi+c$ for the Galileon Lagrangian. The condition (\ref{eq:nopotential})
is also important for another reason. In the Minkowski space-time
($H=0$) the only solution to the equations of motion without matter
corresponds to $\dot{\phi}=0$, provided that $f_{2,\phi}\neq0$.
This implies that the field is frozen, and thus implementing the Vainshtein
mechanism. Moreover the field perturbations would propagate with the
speed of light in the Minkowski background.

The Friedmann equation (\ref{eq:frd1}) can be written in the form
\begin{equation}
\Omega_{r}+\Omega_{m}+\Omega_{{\rm DE}}=1\,,\label{constraint}
\end{equation}
where $\Omega_{r}\equiv\rho_{r}/(3FH^{2})$, $\Omega_{m}\equiv\rho_{m}/(3FH^{2})$,
and 
\begin{equation}
\Omega_{{\rm DE}}\equiv-\frac{\dot{F}}{HF}-\epsilon_{2}\frac{\dot{f}_{2}^{2}}{6FH^{2}}+\frac{\dot{f}_{3}^{3}}{HF}-\epsilon_{4}\frac{15\dot{f}_{4}^{4}}{2F}+\frac{7H\dot{f}_{5}^{5}}{F}-\frac{4H\dot{\xi}}{F}\,.\label{Omecon}
\end{equation}
 In order to realize the late-time dS solutions we take the power-law
functions for $F(\phi)$, $f_{i}(\phi)$, and $\xi(\phi)$ in terms
of $\phi$. We can classify the theories into two classes: (i) $F$
is constant, and (ii) $F$ depends on the field $\phi$.

\subsection{Constant $F$}
\label{conFsec} 

At the dS point ($H=H_{{\rm dS}}=$\,constant) we
require that each term in Eq.~(\ref{Omecon}) does not vary in time.
For constant $F$, the functions $f_{i}$ and $\xi$ need to have
the field dependence proportional to $\phi$. Then the dS solutions
can be realized for constant $\dot{\phi}$. In order to have dimensionless
couplings $d_{i}$ of the orders of unity, we write the functions
$F$, $f_{i}$, and $\xi$ in the forms 
\begin{equation}
F=M_{{\rm pl}}^{2},\qquad f_{2}=d_{2}\phi,\qquad f_{3}=d_{3}\frac{\phi}{M},\qquad f_{4}=d_{4}\frac{\phi}{M^{3/2}},\qquad f_{5}=d_{5}\frac{\phi}{M^{9/5}},\qquad\xi=d_{\xi}\frac{\phi}{M_{\xi}}\,,\label{funchoice}
\end{equation}
 where $M_{{\rm pl}}=2.43\times10^{18}\,$GeV is the reduced Planck
mass. Together with the dimensionless constants $d_{i}$ ($i=2,\cdots,5)$
and $d_{\xi}$ we have introduced the mass scales 
\begin{eqnarray}
 &  & M\equiv\left(\frac{M_{{\rm pl}}}{H_{{\rm dS}}}\right)^{1/3}H_{{\rm dS}}\approx10^{-40}M_{{\rm pl}}\,,\label{eq:muh0}\\
 &  & M_{\xi}\equiv\frac{H_{{\rm dS}}^{2}}{M_{{\rm pl}}}\approx
 10^{-120}M_{{\rm pl}}\,,
\end{eqnarray}
 where we have used $H_{{\rm dS}}\approx10^{-60}M_{{\rm pl}}$.

Defining the mass scales $M$ and $M_{\xi}$ as given above, we can
express the density parameter $\Omega_{{\rm DE}}$ in the form 
\begin{equation}
\Omega_{{\rm DE}}=-\frac{1}{6}\epsilon_{2}d_{2}^{2}x^{2}+d_{3}^{3}xy-\frac{15}{2}\epsilon_{4}d_{4}^{4}y^{2}+7d_{5}^{5}\frac{y^{3}}{x}-4d_{\xi}\frac{y}{x}\,,\label{Omedef}
\end{equation}
 where 
\begin{equation}
x\equiv\frac{\dot{\phi}}{HM_{{\rm pl}}}\,,\qquad y\equiv x^{2}\frac{H^{2}}{H_{{\rm dS}}^{2}}\,.\label{ydef}
\end{equation}
We shall consider the case in which the dimensional variables $x$
and $y$ are not much different from the orders of unity today. In
the asymptotic past we require that $x\ll1$, $y\ll1$, and $y\gg x^{2}$
to recover the General Relativistic behavior. If the coefficients
$d_{i}$ and $d_{\xi}$ are of the orders of unity, the dominant contribution
in Eq.~(\ref{Omedef}) comes from the last term. In such a case,
however, the GB term has been dominant during the whole cosmological
evolution by today. In order to avoid this behavior, we set 
\begin{equation}
d_{\xi}=0\,,
\end{equation}
 when we discuss the cosmological dynamics.

Using Eqs.~(\ref{eq:frd1}) and (\ref{eq:frd2}), we obtain the following relations: 
\begin{eqnarray}
\epsilon_{2}\,(d_{2}x_{{\rm dS}})^{2} & = & {6+9\alpha-12\beta}\,,\label{epre}\\
(d_{3}x_{{\rm dS}})^{3} & = & 2+9\alpha-9\beta\,,\label{d3re}
\end{eqnarray}
 where $x_{{\rm dS}}$ is the value of $x$ at the dS point, and 
\begin{equation}
\alpha\equiv\epsilon_{4}(d_{4}x_{{\rm dS}})^{4}\,,\qquad
\beta \equiv(d_{5}x_{{\rm dS}})^{5}\,.\label{alphadef}
\end{equation}
We note that the theory with the functions (\ref{funchoice}) corresponds
to the covariant Galileon theory discussed in Refs.~\cite{DeffaGal,DT1}.

\subsection{Non-constant $F$}
\label{nonconFsec} 

Let us consider theories in which $F$ depends on the field $\phi$. 
If we take the power-law function of the form
$F\propto\phi^{p}$ with constant $p$, it follows from Eq.~(\ref{Omecon})
that the function $\dot{\phi}/(H\phi)$ is required to be constant
at the dS point. {}From this demand we can restrict the functions
$F$, $f_{i}$, $\xi$ in the forms 
\begin{eqnarray}
&  & F=M_{{\rm pl}}^{2-p}\phi^{p},\qquad f_{2}=d_{2}M_{2}^{1-p/2}\phi^{p/2},\qquad
f_{3}=d_{3}M_{3}^{-p/3}\phi^{p/3},\qquad f_{4}=d_{4}M_{4}^{-1/2-p/4}\phi^{p/4},\nonumber \\
&  & f_{5}=d_{5}M_{5}^{-4/5-p/5}\phi^{p/5},\qquad\xi=d_{\xi}M_{\xi}^{-p}\phi^{p}\,,
\label{funchoice2}
\end{eqnarray}
where $d_{i}$ ($i=2,\cdots,5$), $d_{\xi}$ are dimensionless constants, 
$M_{i}$ ($i=2,\cdots,5$) 
and $M_{\xi}$ are mass scales defined by 
\begin{equation}
M_{2}\equiv M_{{\rm pl}},\quad M_{3}\equiv\left(\frac{M_{{\rm pl}}}{H_{{\rm dS}}}\right)^{(p-2)/p}H_{{\rm dS}},\quad M_{4}\equiv\left(\frac{M_{{\rm pl}}}{H_{{\rm dS}}}\right)^{(p-2)/(p+2)}H_{{\rm dS}},\quad M_{5}\equiv\left(\frac{M_{{\rm pl}}}{H_{{\rm dS}}}\right)^{(p-2)/(p+4)}H_{{\rm dS}},\quad M_{\xi}\equiv M_{3}.
\end{equation}

We note that there are other possibilities to obtain dS solutions, 
depending on the forms of $F(\phi)$.
If we choose the function $F(\phi) \propto e^{\mu \phi}$, where $\mu$
is a constant, then it is possible to realize the dS solution
for the choice $f_2 \propto e^{\mu \phi/2}$, 
$f_3 \propto e^{\mu \phi/3}$, $f_4 \propto e^{\mu \phi/4}$, and
$f_5 \propto e^{\mu \phi/5}$.
In the sense that $\dot{\phi}=$\,constant along the dS solution 
this theory is related with the constant $F$ theory 
given in (\ref{funchoice}).
Since we want to consider the case in which the dS solution 
is realized in a different way, we shall study the cosmological 
dynamics for the theories with (\ref{funchoice2}) in which 
$\dot{\phi}/\phi=$\,constant at the de Sitter solution.

For the theories (\ref{funchoice2}) the Galileon symmetry is explicitly broken. 
However, even for the Galileon action (\ref{funchoice}), 
the Galileon symmetry is restored only on the Minskowski background.
Therefore the Galileon symmetry does not restrict the form of the 
Lagrangian on curved backgrounds. Nonetheless, these theories
may represent an effective action for some more fundamental
theory, e.g., extra-dimensional models. 
In fact the covariant Galileon is the generalization of the 
decoupling limit of the DGP braneworld model.
The existence of dS solutions in this theory 
opens up the possibility of studying 
these generalizations of the original Galileon field.
Moreover the model is not plagued by the Ostrogradski instability
because the field equations remain at second order.
The situation here is not very different from other dark energy
models, such as $f(R)$ gravity, where the Lagrangian is constructed 
by hands to realize the late-time cosmic acceleration but it
is supposed to be originated from some fundamental theory. 
In this directions other papers appeared which tried to generalize 
the Galileon
Lagrangian without imposing the original Galileon 
symmetry \cite{KazuyaGal,Kobayashi1,Kobayashi2,DT1,DPSV}, 
as it is violated, by construction, on curved backgrounds.

The density parameter $\Omega_{{\rm DE}}$ can be expressed as 
\begin{equation}
\Omega_{{\rm DE}}=-p\tilde{x}-\frac{1}{24}\epsilon_{2}d_{2}^{2}p^{2}
\tilde{x}^{2}+\frac{1}{27}d_{3}^{3}p^{3}\tilde{x}\tilde{y}^2
-\frac{15}{512}\epsilon_{4}d_{4}^{4}p^{4}\tilde{y}^{4}+\frac{7}{3125}d_{5}^{5}p^{5}\frac{\tilde{y}^{6}}{\tilde{x}}-4d_{\xi}p\frac{\tilde{y}^2}{\tilde{x}}\,,
\label{OmeDe2}
\end{equation}
where 
\begin{equation}
\tilde{x} \equiv \frac{\dot{\phi}}{H\phi}\,,\qquad
\tilde{y} \equiv \tilde{x} \frac{H}{H_{{\rm dS}}}\,.
\label{tildexy}
\end{equation}
For $p$ and $d_{i}$ ($i=2,\cdots,5$) of the order of unity, we
require that $\tilde{x}\ll1$, $\tilde{y}\ll1$, and 
$\tilde{y} \gg \tilde{x}$ in the asymptotic past. 
Again we shall set $d_{\xi}=0$ in order to
avoid that the last term in Eq.~(\ref{OmeDe2}) always dominates
the cosmological dynamics.

At the dS point we have the following relations 
\begin{eqnarray}
\epsilon_{2}(d_{2}p\tilde{x}_{{\rm dS}})^{2} & = & \frac{24(2p^{2}
\tilde{x}_{{\rm dS}}^{2}+5p\tilde{x}_{{\rm dS}}+9)}{
p\tilde{x}_{{\rm dS}}+9}+\frac{9}{64}\tilde{\alpha}-\frac{48}{3125}\tilde{\beta}\,,\label{gened2} \\
(d_{3}p\tilde{x}_{{\rm dS}})^{3} & = & \frac{81(p\tilde{x}_{{\rm dS}}+2)(p\tilde{x}_{{\rm dS}}+3)}{p\tilde{x}_{{\rm dS}}+9}+\frac{243}{256}\tilde{\alpha}-\frac{243}{3125}\tilde{\beta}\,, 
\label{gened3}
\end{eqnarray}
where 
\begin{equation}
\tilde{\alpha}\equiv\epsilon_{4}(d_{4}p\tilde{x}_{{\rm dS}})^{4}\,,\qquad\tilde{\beta}\equiv(d_{5}p\tilde{x}_{{\rm dS}})^{5}\,.\label{alphadef2}
\end{equation}

In the above two theories
we wish to study 1) whether different Galileon-like actions have
some common feature, and 2) how they differ.

%%%%%%%%%%%%%%%%%%%%%%%%%%%%%%%%%%%%%%%%%%%%%%%
\section{Conditions for the avoidance of ghosts 
and Laplacian instabilities}
\label{secins}
%%%%%%%%%%%%%%%%%%%%%%%%%%%%%%%%%%%%%%%%%%%%%%%

In order to discuss the stability of theories described by the Lagrangian
(\ref{Lag}) in the cosmological context, it is necessary to study
linear perturbation theory on the flat FLRW background. Let us consider
the perturbed metric 
\begin{equation}
\de s^{2}=-[1+2\Psi(t,\bm{x})]\,\de t^{2}+\partial_{i}\chi(t,\bm{x})\,\de t\,
\de x^{i}+a^{2}(t)[1+2\Phi(t,\bm{x})]\,\de\bm{x}^{2}\,,\label{eq:perto1}
\end{equation}
 where $\Psi$, $\Phi$, and $\chi$ are scalar metric perturbations.
We have chosen the gauge $\delta\phi=0$ without a non-diagonal scalar
perturbation in the spatial part of the metric, i.e. $\partial_{ij}\gamma=0$
(see Refs.~\cite{Bardeen} for the details of gauge-invariant cosmological
perturbation theory). Taking into account two perfect fluids with
the equations of state $w_{i}=P_{i}/\rho_{i}$ ($i=1,2$), there are
three propagating scalar degrees of freedom. The velocity potentials
$v_{i}$ ($i=1,2$) are related with the energy-momentum tensor ${T_{j}^{0}}^{(i)}$,
as ${T_{j}^{0}}^{(i)}=-(\rho_{i}+P_{i})\partial_{j}v_{i}$ ($i=1,2$).

Expanding the action (\ref{action}) at second-order in perturbations,
we find that the field $\Psi$ can be integrated out together with
$\chi$. Introducing the vector $\vec{{\cal Q}}=(v_{1},v_{2},\Phi)$,
we obtain the following second-order action for scalar perturbations
(see Refs.~\cite{DeFelice,DMT} for the details of such analysis):
\begin{equation}
\delta S_{S}^{(2)}=\frac{1}{2}\int\de t\,\de^{3}x\, a^{3}\,
\left[\dot{\vec{{\cal Q}}}^{t}\bm{A}\dot{\vec{{\cal Q}}}
-\frac{1}{a^{2}}\nabla{\vec{{\cal Q}}}^{t}\,\bm{C}\,
\nabla{\vec{{\cal Q}}}-\dot{\vec{{\cal Q}}}^{t}\bm{B}{\vec{{\cal Q}}}
-{\vec{{\cal Q}}}^{t}\bm{D}{\vec{{\cal Q}}}\,\right]\,,\label{eq:perto2}
\end{equation}
 where $\bm{A}$, $\bm{C}$ and $\bm{D}$ are $3\times3$ symmetric
matrices and $\bm{B}$ is an antisymmetric matrix (we do not write
explicit forms for them).

Let us consider tensor perturbations with $\delta g_{ij}=a^{2}h_{ij}$,
where $h_{ij}$ is traceless ($h^{i}{}_{i}=0$) and divergence-free
($h^{ij}{}_{,j}=0$). We also expand the action (\ref{action}) at
second-order in terms of the two polarization modes, 
$h_{ij}=h_{\oplus}\,\epsilon_{ij}^{\oplus}+h_{\otimes}\,\epsilon_{ij}^{\otimes}$,
where $\epsilon_{ij}^{\oplus}$ and $\epsilon_{ij}^{\otimes}$ are
the polarization tensors. For the polarization mode $h_{\oplus}$,
the perturbed action is given by 
\begin{equation}
\delta S_{T}^{(2)}=\frac{1}{2}\int\de t\,\de^{3}x\, a^{3}\, 
Q_{T}\left[\dot{h}_{\oplus}^{2}-\frac{c_{T}^{2}}{a^{2}}\,
(\nabla h_{\oplus})^{2}\right]\,,\label{eq:parto2GW}
\end{equation}
 where we will show the explicit forms of $Q_{T}$ and $c_{T}^{2}$
later. The conditions for the avoidance of ghosts and Laplacian instabilities
of tensor perturbations correspond to $Q_{T}>0$ and $c_{T}^{2}>0$,
respectively. Note that the same expression also holds for $h_{\otimes}$.

In Secs.~\ref{subghost} and \ref{sublap} we study the general theories
described by the Lagrangian (\ref{Lag}) without imposing that $f_{1}$
and $\xi$ are zero. In Sec.~\ref{subdS} we shall apply our formula
to specific theories with $f_{1}=0$ and $\xi=0$.

\subsection{No-ghost conditions}

\label{subghost} In order to avoid that a ghost mode appears for
scalar perturbations, the matrix $\bm{A}$ needs to be positive definite.
This leads to the following three no-ghost conditions 
\begin{eqnarray}
&  & \frac{1+w_{1}}{w_{1}}\,\rho_{1}>0\,,
\label{eq:constro1}\\
&  & \frac{1+w_{2}}{w_{2}}\,\rho_{2}>0\,,
\label{eq:constro2}\\
&  & Q_{S}\equiv\frac{\gamma_{1}}{9}
\frac{4(\rho_{1}+\rho_{2}+\gamma_{2})\gamma_{1}
-9\gamma_{3}^{2}}{\gamma_{3}^{2}}>0\,,
\label{eq:genQs}
\end{eqnarray}
where 
\begin{eqnarray}
 & \gamma_{1} & \equiv-6F-9\epsilon_{4}\dot{f}_{4}^{4}+18H\dot{f}_{5}^{5}-24H\dot{\xi}\,,\\
 & \gamma_{2} & \equiv-9FH^{2}-9H\dot{F}-f_{1}/2-3\epsilon_{2}\dot{f}_{2}^{2}/2+15H\dot{f}_{3}^{3}-315\epsilon_{4}H^{2}\dot{f}_{4}^{4}/2+189H^{3}\dot{f}_{5}^{5}-60H^{3}\dot{\xi}\,,\\
 & \gamma_{3} & \equiv-4FH-2\dot{F}+2\dot{f}_{3}^{3}-30\epsilon_{4}H\dot{f}_{4}^{4}+42H^{2}\dot{f}_{5}^{5}-24H^{2}\dot{\xi}\,.
\end{eqnarray}
 For radiation ($w_{1}=1/3$) and non-relativistic matter ($w_{2}\simeq+0$)
the conditions (\ref{eq:constro1}) and (\ref{eq:constro2}) are automatically
satisfied. Hence we only need to consider the condition (\ref{eq:genQs})
to avoid the appearance of ghosts for scalar perturbations.

The no-ghost condition for tensor perturbations is given by 
\begin{equation}
Q_{T}=\frac{1}{2}F+\frac{3}{4}\epsilon_{4}\dot{f}_{4}^{4}
-\frac{3}{2}H\dot{f}_{5}^{5}+2H\dot{\xi}>0\,.\label{eq:genQW}
\end{equation}
\subsection{Conditions for the avoidance of Laplacian instabilities}
\label{sublap} 

The propagation speeds $c_{S}$ of three scalar degrees
of freedom is known by solving the equation 
\begin{equation}
\det(c_{S}^{2}\bm{A}-\bm{C})=0\,.\label{eq:cc}
\end{equation}
 The standard velocities for two perfect fluids correspond to $c_{S}^{2}=w_{1}$
and $c_{S}^{2}=w_{2}$, which are positive for both radiation and
non-relativistic matter. The stability condition coming from the third
solution is given by 
\begin{equation}
c_{S}^{2}=\frac{2\gamma_{1}^{2}\,[\dot{\gamma}_{3}-2(1+w_{1})\rho_{1}-2(1+w_{2})\rho_{2}-\gamma_{3}H]-4\dot{\gamma}_{1}\gamma_{1}\gamma_{3}+\gamma_{3}^{2}\gamma_{4}}{\gamma_{1}[4(\rho_{1}+\rho_{2}+\gamma_{2})\gamma_{1}-9\gamma_{3}^{2}]}>0\,,\label{cS2}
\end{equation}
 where 
\begin{equation}
\gamma_{4}\equiv-18F+9\epsilon_{4}\dot{f}_{4}^{4}+54\dot{f}_{5}^{4}\ddot{f}_{5}-72\ddot{\xi}\,.\label{eq:alp5}
\end{equation}
 Under the no-ghost condition (\ref{eq:genQs}), this translates to
\begin{equation}
2\gamma_{1}^{2}\,[\dot{\gamma}_{3}-2(1+w_{1})\rho_{1}-2(1+w_{2})\rho_{2}-\gamma_{3}H]-4\dot{\gamma}_{1}\gamma_{1}\gamma_{3}+\gamma_{3}^{2}\gamma_{4}>0\,.\label{noinst}
\end{equation}
The Laplacian instability of tensor perturbations is absent provided
that the propagation speed squared is positive: 
\begin{equation}
c_{T}^{2}=\frac{2F-\epsilon_{4}\dot{f}_{4}^{4}-6\dot{f}_{5}^{4}\ddot{f}_{5}
+8\ddot{\xi}}{2F+3\epsilon_{4}\dot{f}_{4}^{4}-6H\dot{f}_{5}^{5}
+8H\dot{\xi}}>0\,.
\label{eq:noinstGW}
\end{equation}

\subsection{de Sitter stability}

\label{subdS} On the dS background there are no matter fields, so
that only one scalar mode propagates. The second-order action for
scalar perturbations at the dS fixed point is 
\begin{equation}
\delta S_{S}^{(2)}=\frac{1}{2}\int\de t\,\de^{3}x\, a^{3}\, Q_{S}\left[\dot{\Phi}^{2}-\frac{c_{S}^{2}}{a^{2}}\,(\nabla\Phi)^{2}\right]\,,\label{eq:actdeS}
\end{equation}
 where $Q_{S}$ and $c_{S}^{2}$ correspond to those given in Eqs.~(\ref{eq:genQs})
and (\ref{cS2}) with the limits $H\to H_{{\rm dS}}$ and $\rho_{1,2}\to0$.
Let us discuss the conditions for the avoidance of ghosts and instabilities
on the dS solutions we have discussed in Sec.~\ref{cosmosec}. We
shall consider two theories described by the functions (\ref{funchoice})
and (\ref{funchoice2}), with $f_{1}=0$ and $\xi=0$.

\subsubsection{Constant $F$}

\label{confsec} For the theory we discussed in Sec.~\ref{conFsec},
the conditions (\ref{eq:genQs}), (\ref{cS2}), (\ref{eq:genQW}),
and (\ref{eq:noinstGW}) reduce to 
\begin{eqnarray}
\frac{Q_{S}}{M_{{\rm pl}}^{2}} & = & \frac{4-9(\alpha-2\beta)^{2}}{3(\alpha-2\beta)^{2}}>0\,,\label{eq:qcs}\\
c_{S}^{2} & = & \frac{(\alpha-2\beta)(4+15\alpha^{2}-48\alpha\beta+36\beta^{2})}{2[4-9(\alpha-2\beta)^{2}]}>0\,,\label{csds}\\
\frac{Q_{T}}{M_{{\rm pl}}^{2}} & = & \frac{1}{4}\,(2+3\alpha-6\beta)>0\,,\label{eq:qcs3}\\
c_{T}^{2} & = & \frac{2-\alpha}{2+3\alpha-6\beta}>0\,.\label{eq:qcs4}
\end{eqnarray}
 In Sec.~\ref{dynamicssec} we will show the allowed parameter space
in the $(\alpha,\beta)$ plane after deriving other conditions.

{}From the action (\ref{eq:actdeS}) we obtain the equation for $\Phi$
in Fourier space: 
\begin{equation}
\frac{1}{a^{3}Q_{S}}\frac{{\rm d}}{{\rm d}t}
\left(a^{3}Q_{S}\dot{\Phi}\right)+c_{S}^{2}\frac{k^{2}}{a^{2}}\Phi=0\,,
\end{equation}
where $k$ is a comoving wavenumber. The solution for the homogeneous
perturbation ($k=0$) is 
\begin{equation}
\Phi=c_{1}+c_{2}\int\frac{1}{a^{3}Q_{S}}{\rm d}t\,,\label{Phiso}
\end{equation}
 where $c_{1}$ and $c_{2}$ are integration constants. Since $Q_{S}$
is constant on the dS solution, the second term on the right hand
side of Eq.~(\ref{Phiso}) decays with time by noting that the scale
factor evolves as $a\propto e^{H_{{\rm dS}}t}$. For the same reason,
the tensor perturbation remains always stable in the limit $k\to0$.
This means that the dS fixed point is always classically stable under
homogeneous perturbations.

\subsubsection{Non-constant $F$}

For the theory we discussed in Sec.~\ref{nonconFsec}, the conditions
(\ref{eq:genQW}) and (\ref{eq:qcs4}) reduce to 
\begin{eqnarray}
\frac{Q_{T}}{M_{{\rm pl}}^{2}} & = & \left(\frac{\phi}{M_{{\rm pl}}}\right)^{p}\left[\frac{1}{2}+\frac{3}{1024}\tilde{\alpha}-\frac{3}{6250}\tilde{\beta}\right]>0\,,\\
c_{T}^{2} & = & 1-\frac{4[15625\tilde{\alpha}+384\tilde{\beta}(p\tilde{x}_{{\rm dS}}-5)]}{15625(3\tilde{\alpha}+512)-7680\tilde{\beta}}>0\,.
\end{eqnarray}
The expressions for $Q_{S}$ and $c_{S}^{2}$ are more involved,
but $Q_{S}$ is proportional to $\phi^{p}$ as in the case of $Q_{T}$.
Integrating the relation $\dot{\phi}/(H\phi)=\tilde{x}_{{\rm dS}}={\rm constant}$,
it follows that $\phi\propto a^{\tilde{x}_{{\rm dS}}}$. Since $a^{3}Q_{s}\propto e^{(3+p\tilde{x}_{{\rm dS}})H_{{\rm dS}}t}$,
we find from Eq.~(\ref{Phiso}) that the homogenous perturbation
$\Phi$ evolves as 
\begin{equation}
\Phi=\tilde{c}_{1}+\tilde{c}_{2}e^{-(3+p\tilde{x}_{{\rm dS}})H_{{\rm dS}}t}\,,
\end{equation}
where $\tilde{c}_{1}$ and $\tilde{c}_{2}$ are constants. Hence
the dS point is classically stable for 
\begin{equation}
3+p\tilde{x}_{{\rm dS}}>0\,.
\label{dsstability}
\end{equation}
The stability condition (\ref{dsstability}) is satisfied for 
$|p\tilde{x}_{{\rm dS}}|\ll1$.
In this regime the conditions (\ref{eq:genQs}) and (\ref{cS2}) for
the scalar perturbation reduce to 
\begin{eqnarray}
\frac{Q_{S}}{M_{{\rm pl}}^{2}} & = & 
\left(\frac{\phi}{M_{{\rm pl}}}\right)^{p}
\left[\frac{243}{(p\tilde{x}_{{\rm dS}})^{2}}
+{\cal O}(\tilde{x}_{{\rm dS}}^{-1})\right]>0\,,
\label{eq:Qc2sx}\\
c_{S}^{2} & = & -\frac{p\tilde{x}_{{\rm dS}}}{27}
+{\cal O}(\tilde{x}_{{\rm dS}}^{2})>0\,.
\label{eq:Cc2sx}
\end{eqnarray}
For positive $\phi$ the no-ghost condition (\ref{eq:Qc2sx}) is
satisfied. If $p\tilde{x}_{{\rm dS}}<0$, the Laplacian instability
of the scalar mode can be avoided.

%%%%%%%%%%%%%%%%%%%%%%%%%%%%%%%%%%%%%%%%%%%%%%%%%%%%%%%%%%%%
\section{Cosmology based on the covariant Galileon theory}
\label{dynamicssec} 
%%%%%%%%%%%%%%%%%%%%%%%%%%%%%%%%%%%%%%%%%%%%%%%%%%%%%%%%%%%%

First we study cosmological dynamics for the covariant 
Galileon theory described by the functions (\ref{funchoice})
with $d_{\xi}=0$. This was partially discussed in the letter \cite{DT2}, 
but in this paper we shall thoroughly study the cosmology in such a theory
with detailed numerical simulations.

In the presence of radiation ($\rho_{1}=\rho_{r},w_{1}=1/3$)
and non-relativistic matter ($\rho_{2}=\rho_{m},w_{2}\simeq+0$),
we obtain the background equations from Eqs.~(\ref{eq:frd1}) and
(\ref{eq:frd2}): 
\begin{eqnarray}
 &  & 3M_{{\rm pl}}^{2}H^{2}=\rho_{{\rm DE}}+\rho_{r}+\rho_{m}\,,\label{be1}\\
 &  & 3M_{{\rm pl}}^{2}H^{2}+2M_{{\rm pl}}^{2}\dot{H}=-P_{{\rm DE}}-\rho_{r}/3\,,\label{be2}
\end{eqnarray}
 where 
\begin{eqnarray}
\hspace{-0.7cm} &  & \rho_{{\rm DE}}=-\epsilon_{2}d_{2}^{2}\dot{\phi}^{2}/2+3d_{3}^{3}H\dot{\phi}^{3}/M^{3}-45\epsilon_{4}d_{4}^{4}H^{2}\dot{\phi}^{4}/(2M^{6})+21d_{5}^{5}H^{3}\dot{\phi}^{5}/M^{9}\,,\\
\hspace{-0.7cm} &  & P_{{\rm DE}}=-\epsilon_{2}d_{2}^{2}\dot{\phi}^{2}/2-d_{3}^{3}\dot{\phi}^{2}\ddot{\phi}/M^{3}+3\epsilon_{4}d_{4}^{4}\dot{\phi}^{3}[8H\ddot{\phi}+(3H^{2}+2\dot{H})\dot{\phi}]/(2M^{6})-3d_{5}^{5}H\dot{\phi}^{4}[5H\ddot{\phi}+2(H^{2}+\dot{H})\dot{\phi}]/M^{9}\,.
\end{eqnarray}
 The continuity equations for radiation and non-relativistic matter
are given, respectively, by 
\begin{equation}
\dot{\rho}_{r}+4H\rho_{r}=0\,,\qquad\dot{\rho}_{m}+3H\rho_{m}=0\,.
\label{be3}
\end{equation}
 {}From Eqs.~(\ref{be1}), (\ref{be2}), and (\ref{be3}) the dark
component also obeys the continuity equation 
\begin{equation}
\dot{\rho}_{{\rm DE}}+3H(\rho_{{\rm DE}}+P_{{\rm DE}})=0\,.\label{rhoDE}
\end{equation}
 We define the dark energy equation of state $w_{{\rm DE}}$ and the
effective equation of state $w_{{\rm eff}}$, as 
\begin{equation}
w_{{\rm DE}}\equiv\frac{P_{{\rm DE}}}{\rho_{{\rm DE}}}\,,\qquad w_{{\rm eff}}\equiv-1-\frac{2\dot{H}}{3H^{2}}\,,
\end{equation}
 where the latter is known by the background expansion history of
the Universe. Using Eq.~(\ref{rhoDE}) together with the relation
$\rho_{{\rm DE}}=3M_{{\rm pl}}^{2}H^{2}\Omega_{{\rm DE}}$, it follows
that 
\begin{equation}
w_{{\rm DE}}=w_{{\rm eff}}-\frac{\Omega_{{\rm DE}}'}{3\Omega_{{\rm DE}}}\,,
\end{equation}
 where a prime represents a derivative with respect to $N=\ln a$.

Each term in Eq.~(\ref{Omedef}) has the difference of the order
of $x/y$. The highest-order term in $\Omega_{{\rm DE}}$ comes from
the term ${\cal L}_{5}$, i.e., of the order of $y^{3}/x$. For the
dynamical analysis given below, it is convenient to introduce the
following quantities 
\begin{equation}
r_{1}\equiv\frac{xx_{{\rm dS}}}{y}=\frac{x_{{\rm dS}}}{x}
\left(\frac{H_{{\rm dS}}}{H}\right)^{2}\,,\qquad r_{2}\equiv
\frac{y^{3}}{xx_{{\rm dS}}^{5}}=\left(\frac{x}{x_{{\rm dS}}}
\right)^{2}\frac{1}{r_{1}^{3}}\,.\label{eq:r1r2}
\end{equation}
 At the dS fixed point one has $r_{1}=1$ and $r_{2}=1$. In terms
of $r_{1}$ and $r_{2}$ the density parameter (\ref{Omecon}) can
be written as 
\begin{equation}
\Omega_{{\rm DE}}=-\frac{1}{2}(3\alpha-4\beta+2)
r_{1}^{3}r_{2}+(9\alpha-9\beta+2)r_{1}^{2}r_{2}
-\frac{15}{2}\alpha r_{1}r_{2}+7\beta r_{2}\,,\label{OmeDE}
\end{equation}
 where $\alpha$ and $\beta$ are defined in Eq.~(\ref{alphadef}).
Here we have employed the relations (\ref{epre}) and (\ref{d3re})
to eliminate the terms $\epsilon_{2}d_{2}^{2}$ and $d_{3}^{3}$.

It is convenient to use the variables $\alpha$ and $\beta$ for several
reasons. First, the coefficients of physical quantities {[}such as
$\Omega_{{\rm DE}}$ in Eq.~(\ref{OmeDE}){]}, autonomous equations,
quantities related with no-ghost and stability conditions can be expressed
in terms of $\alpha$ and $\beta$. Second, the equations of motion,
together with linear perturbation theory, are not subject to change
under the following change of parameters $x_{{\rm dS}}\to\gamma x_{{\rm dS}}$
and $d_{i}\to d_{i}/\gamma$ (with $i=2,3,4,5$), where $\gamma$
is a real number. In this case, depending on the parameter $\gamma$,
there are infinite choices for the Lagrangian coefficients $d_{i}$
that lead to the identical physics for the same $\alpha$ and $\beta$.
Therefore, constraining the parameter space in terms of $\alpha$
and $\beta$ allows us to remove the arbitrariness of the $\gamma$
rescaling. This also shows that one can set $x_{{\rm dS}}=1$ without
losing generalities.

If $r_{1}\ll1$ at early times, the highest-order term ${\cal L}_{5}$
gives the dominant contribution to the dark energy density $\Omega_{{\rm DE}}$.
In this case it is expected that the cosmological Vainshtein mechanism
can be at work to recover the General Relativistic behavior. If $r_{1}\gg1$
initially, the dominant contribution to $\Omega_{{\rm DE}}$ comes
from the term ${\cal L}_{2}$. In this case the field energy density
decreases rapidly as in the standard massless scalar field and hence
the solutions do not approach the dS fixed point at late times.

The conditions (\ref{eq:genQs}), (\ref{eq:genQW}), (\ref{cS2}),
and (\ref{eq:noinstGW}) for the avoidance of ghosts and instabilities
of scalar and tensor perturbations reduce to 
\begin{eqnarray}
 &  & \frac{Q_{S}}{M_{{\rm pl}}^{2}}=-\frac{6(1+\mu_{1})(\mu_{1}+\mu_{2}+\mu_{1}\mu_{2}-2\mu_{3}-\mu_{3}^{2})}{(1+\mu_{3})^{2}}>0\,,\label{st1}\\
 &  & \frac{Q_{T}}{M_{{\rm pl}}^{2}}=\frac{1}{2}+\frac{3}{4}\alpha r_{1}r_{2}-\frac{3}{2}\beta r_{2}>0\,,\\
 &  & c_{S}^{2}=\frac{(1+\mu_{1})^{2}[2\mu_{3}'-(1+\mu_{3})(5+3w_{{\rm eff}})+4\Omega_{r}+3\Omega_{m}]-4\mu_{1}'(1+\mu_{1})(1+\mu_{2})+2(1+\mu_{3})^{2}(1+\mu_{4})}{6(1+\mu_{1})(\mu_{1}+\mu_{2}+\mu_{1}\mu_{2}-2\mu_{3}-\mu_{3}^{2})}>0\,,\\
 &  & c_{T}^{2}=\frac{2r_{1}(2-\alpha r_{1}r_{2})-3\beta(r_{2}r_{1}'+r_{1}r_{2}')}{2r_{1}(2+3\alpha r_{1}r_{2}-6\beta r_{2})}>0\,,\label{st4}
\end{eqnarray}
 where 
\begin{eqnarray}
 &  & \mu_{1}\equiv3\alpha r_{1}r_{2}/2-3\beta r_{2}\,,\\
 &  & \mu_{2}\equiv(3\alpha-4\beta+2)r_{1}^{3}r_{2}/2-2(9\alpha-9\beta+2)r_{1}^{2}r_{2}+45\alpha r_{1}r_{2}/2-28\beta r_{2}\,,\\
 &  & \mu_{3}\equiv-(9\alpha-9\beta+2)r_{1}^{2}r_{2}/2+15\alpha r_{1}r_{2}/2-21\beta r_{2}/2\,,\\
 &  & \mu_{4}\equiv-\alpha r_{1}r_{2}/2-3\beta r_{2}(r_{1}'/r_{1}+r_{2}'/r_{2})/4\,.
\end{eqnarray}
 {}From Eqs.~(\ref{eq:frd1}), (\ref{eq:frd2}), and (\ref{conser})
we obtain the following differential equations for the variables $r_{1}$,
$r_{2}$, and $\Omega_{r}$: 
\begin{eqnarray}
\hspace{-0.7cm}r_{1}' & = & \frac{1}{\Delta}\left(r_{1}-1\right)r_{1}\left[r_{1}\left(r_{1}(-3\alpha+4\beta-2)+6\alpha-5\beta\right)-5\beta\right]\nonumber \\
 &  & {}\times\left[2\left(\Omega_{r}+9\right)+3r_{2}\left(r_{1}^{3}
 (-3\alpha+4\beta-2)+2r_{1}^{2}(9\alpha-9\beta+2)-15r_{1}\alpha+14\beta\right)\right]\,, \label{eq:DRr1} \\
\hspace{-0.7cm}r_{2}' & = & -\frac{1}{\Delta}[r_{2}(6r_{1}^{2}(r_{2}(45\alpha^{2}-4(9\alpha+2)\beta+36\beta^{2})-(\Omega_{r}-7)(9\alpha-9\beta+2))+r_{1}^{3}(-2(\Omega_{r}+33)(3\alpha-4\beta+2)\nonumber \\
 &  & -3r_{2}(-2(201\alpha+89)\beta+15\alpha(9\alpha+2)+356\beta^{2}))-3r_{1}\alpha(-28\Omega_{r}+123r_{2}\beta+36)+10\beta(-11\Omega_{r}+21r_{2}\beta-3)\nonumber \\
 &  & +3r_{1}^{4}r_{2}(9\alpha^{2}-30\alpha(4\beta+1)+2(2-9\beta)^{2})+3r_{1}^{6}r_{2}(3\alpha-4\beta+2)^{2}+3r_{1}^{5}r_{2}(9\alpha-9\beta+2)(3\alpha-4\beta+2))], \label{eq:DRr2} \\
\hspace{-0.7cm}\Omega_{r}' & = & \frac{2}{\Delta}\Omega_{r}[r_{1}^{2}(4(\Omega_{r}-1)(9\alpha-9\beta+2)+6r_{2}(-15\alpha^{2}+36\alpha\beta+4(2-9\beta)\beta))-2r_{1}^{3}((\Omega_{r}-1)(3\alpha-4\beta+2)\nonumber \\
 &  & +9r_{2}(18(\alpha+1)\beta+\alpha(9\alpha+2)-36\beta^{2}))+12r_{1}\alpha(-3\Omega_{r}+22r_{2}\beta+3)-10\beta(-4\Omega_{r}+21r_{2}\beta+4)\nonumber \\
 &  & +r_{1}^{4}r_{2}(549\alpha^{2}+\alpha(330-840\beta)+2(2-9\beta)^{2})+3r_{1}^{6}r_{2}(3\alpha-4\beta+2)^{2}-12r_{1}^{5}r_{2}(9\alpha-9\beta+2)(3\alpha-4\beta+2)], \label{eq:DRr3}
\end{eqnarray}
where 
\begin{eqnarray}
\Delta & \equiv & 2r_{1}^{4}r_{2}[72\alpha^{2}+30\alpha(1-5\beta)+(2-9\beta)^{2}]+4r_{1}^{2}[9r_{2}(5\alpha^{2}+9\alpha\beta+(2-9\beta)\beta)+2(9\alpha-9\beta+2)]\nonumber \\
 &  & +4r_{1}^{3}[-3r_{2}\left(-2(15\alpha+1)\beta+3\alpha(9\alpha+2)+4\beta^{2}\right)-3\alpha+4\beta-2]-24r_{1}\alpha(16r_{2}\beta+3)+10\beta(21r_{2}\beta+8)\,.
\end{eqnarray}
 The Hubble parameter obeys the following equation 
\begin{equation}
\frac{H'}{H}=-\frac{5r_{1}'}{4r_{1}}-\frac{r_{2}'}{4r_{2}}\,,
\end{equation}
where $r_{1}'/r_{1}$ and $r_{2}'/r_{2}$ are known from Eqs.~(\ref{eq:DRr1})
and (\ref{eq:DRr2}).

\subsection{Tracker solutions ($r_{1}=1$)}

{}From Eq.~(\ref{eq:DRr1}) we find that there is an equilibrium
point characterized by 
\begin{equation}
r_{1}=1\,,\label{eq:soldR1}
\end{equation}
 at which the density parameter (\ref{OmeDE}) reduces to 
\begin{equation}
\Omega_{{\rm DE}}=r_{2}\,.\label{OmeDEre}
\end{equation}
 {}From Eq.~(\ref{eq:r1r2}) we find that $xH^{2}=$\,constant
along the solution (\ref{eq:soldR1}). Hence the field velocity evolves
as 
\begin{equation}
\dot{\phi}\propto H^{-1}\,,
\end{equation}
 which has the dependence $\dot{\phi}\propto t$ during the radiation
and matter eras. Since the field is effectively frozen at early times,
this shows the implementation of the cosmological Vainshtein mechanism.

Along the solution (\ref{eq:soldR1}), the other two equations can
be written as follows 
\begin{eqnarray}
r_{2}' & = & \frac{2r_{2}\left(3-3r_{2}+\Omega_{r}\right)}{1+r_{2}}\,,\label{eq:dR2}\\
\Omega_{r}' & = & \frac{\Omega_{r}\left(\Omega_{r}-1-7r_{2}\right)}{1+r_{2}}\,,\label{eq:dOmr}
\end{eqnarray}
 which do not depend on $\alpha$ and $\beta$. We then have the following
three fixed points 
\begin{equation}
{\rm (A)}~(r_{1},r_{2},\Omega_{r})=(1,0,1)\,,\qquad{\rm (B)}~(r_{1},r_{2},\Omega_{r})=(1,0,0)\,,\qquad{\rm (C)}~(r_{1},r_{2},\Omega_{r})=(1,1,0)\,.
\end{equation}
 The points (A) and (B) can be realized during the radiation and matter
eras, respectively, whereas the point (C) corresponds to the dS solution.

The stabilities of these fixed points can be analyzed by considering
linear perturbations $\delta r_{1}$, $\delta r_{2}$, and $\delta\Omega_{r}$
about them. For example, the perturbation $\delta r_{1}$ satisfies
\begin{equation}
\delta r_{1}'=-\frac{9+\Omega_{r}+3r_{2}}{2(1+r_{2})}\delta r_{1}\,.
\end{equation}
 This shows that, in the regime $0\leq r_{2}\leq1$ and $\Omega_{r}\geq0$,
the solution is stable in the direction of $r_{1}$. Defining the
vector $\delta\bm{r}={}^{t}(\delta r_{1},\delta r_{2},\delta r_{3})$,
one can write the perturbation equations in the form 
\begin{equation}
\delta\bm{r}'={\cal M}\,\delta\bm{r}\,,
\end{equation}
 where ${\cal M}$ is the $3\times3$ matrix. The eigenvalues of the
matrix ${\cal M}$ for the points (A), (B), (C) are given by 
\begin{equation}
{\rm (A)}~(8,1,-5)\,,\qquad{\rm (B)}~(6,-1,-9/2)\,,\qquad{\rm (C)}~(-3,-3,-4)\,.
\end{equation}
 This shows that (A) and (B) are saddle, while (C) is stable. Hence
the solutions finally approach the stable dS point (C). This dS stability
is consistent with the analysis in Sec.~\ref{confsec} based on homogeneous
perturbations. The solution (\ref{eq:soldR1}) can be regarded as
a tracker that attracts solutions with different initial conditions
to a common trajectory.

Along the tracker we have $\rho_{{\rm DE}}=3M^{6}/H^{2}$, 
$P_{{\rm DE}}=-3M^{6}(2+w_{{\rm eff}})/H^{2}$,
and 
\begin{equation}
w_{{\rm DE}}=-2-w_{{\rm eff}}=-\frac{\Omega_{r}+6}{3(r_{2}+1)}\,,
\qquad w_{{\rm eff}}=\frac{\Omega_{r}-6r_{2}}{3(r_{2}+1)}\,.\label{wdetra}
\end{equation}
 During the cosmological sequence of radiation, matter, and dS eras
the dark energy equation of state evolves as $w_{{\rm DE}}=-7/3\to-2\to-1$,
whereas the evolution of the effective equation of state is $w_{{\rm eff}}=1/3\to0\to-1$.
This peculiar evolution of $w_{{\rm DE}}$ can be useful to constrain
the covariant Galileon theory from observations.

Equations (\ref{eq:dR2}) and (\ref{eq:dOmr}) are simple enough to
be solved analytically. In fact, combining Eqs.~(\ref{eq:dR2}) and
(\ref{eq:dOmr}), it follows that 
\begin{equation}
\frac{r_{2}'}{r_{2}}=8+2\frac{\Omega_{r}'}{\Omega_{r}}\,,\label{eq:dert1}
\end{equation}
 which has the solution 
\begin{equation}
r_{2}=c_{1}a^{8}\Omega_{r}^{2}\,,\label{eq:solR2}
\end{equation}
 where $c_{1}$ is a constant of integration. Substituting this solution
into Eq.~(\ref{eq:dOmr}), we find two branches that differ from
each other in the early cosmological limit. The viable branch of solutions
is given by 
\begin{equation}
\Omega_{r}={\frac{c_{2}a-1+\sqrt{1-2c_{2}a+{c_{2}}^{2}{a}^{2}
+4c_{1}{a}^{8}}}{2c_{1}{a}^{8}}}\,,\label{eq:omsol}
\end{equation}
 where $c_{2}$ is another constant. Since $\Omega_{r}\simeq1+c_{2}\, a$
at early times ($a\ll1$), we require that $c_{2}<0$ (provided $\Omega_{{\rm DE}}>0$).

The coefficients $c_{1}$ and $c_{2}$ can be found by using the present
density parameters of radiation and non-relativistic matter, i.e.
$\Omega_{r}(a=1)=\Omega_{r}^{(0)}$ and $\Omega_{m}(a=1)=\Omega_{m}^{(0)}$.
Using the relation (\ref{OmeDEre}) as well, we find 
\begin{eqnarray}
c_{1}=\frac{1-\Omega_{m}^{(0)}-\Omega_{r}^{(0)}}{(\Omega_{r}^{(0)})^{2}}\,,\qquad c_{2}=-\frac{\Omega_{m}^{(0)}}{\Omega_{r}^{(0)}}\,.\label{eq:eqs1}
\end{eqnarray}
 The density parameter of dark energy evolves as 
\begin{equation}
\Omega_{{\rm DE}}=c_{1}a^{8}\Omega_{r}^{2}\,.\label{eq:omede}
\end{equation}
 Hence the density parameters $\Omega_{{\rm DE}}$, $\Omega_{r}$,
and $\Omega_{m}$ as well as $w_{{\rm DE}}$ and $w_{{\rm eff}}$
are analytically known in terms of the function of $a$ 
(or the redshift $z=1/a-1$).

At the dS point (C) the conditions for the avoidance of ghosts and
instabilities have been already estimated in Eqs.~(\ref{eq:qcs})-(\ref{eq:qcs4}).
Let us consider the points (A) and (B), which are characterized by
$r_{1}=1$ and $r_{2}\ll1$. In this case Eqs.~(\ref{st1})-(\ref{st4})
are simplified to give 
\begin{eqnarray}
 &  & Q_{S}/M_{{\rm pl}}^{2}\simeq3(2-3\alpha+6\beta)r_{2}>0\,,\label{at1}\\
 &  & Q_{T}/M_{{\rm pl}}^{2}=1/2+3(\alpha-2\beta)r_{2}/4>0\,,\label{at2}\\
 &  & c_{S}^{2}\simeq\frac{8+10\alpha-9\beta+\Omega_{r}(2+3\alpha-3\beta)}
 {3(2-3\alpha+6\beta)}>0\,,\label{at3}\\
 &  & c_{T}^{2}\simeq1-(4\alpha+3\beta+3\beta\Omega_{r})r_{2}/2>0\,.\label{at4}
\end{eqnarray}
 Since $r_{2}\ll1$ the conditions (\ref{at2}) and (\ref{at4}) are
automatically satisfied. {}From Eq.~(\ref{at1}) the sign change
of $r_{2}$ means the appearance of the scalar ghost. If we choose
the initial conditions with $r_{2}>0$, then Eq.~(\ref{at1}) requires
that 
\begin{equation}
2-3\alpha+6\beta>0\,.\label{at1d}
\end{equation}

%%%%%%%%%%%%%%%%%%%%%%%%%%%%%
\begin{figure}
\begin{centering}
\includegraphics[width=3.3in,height=3.4in]{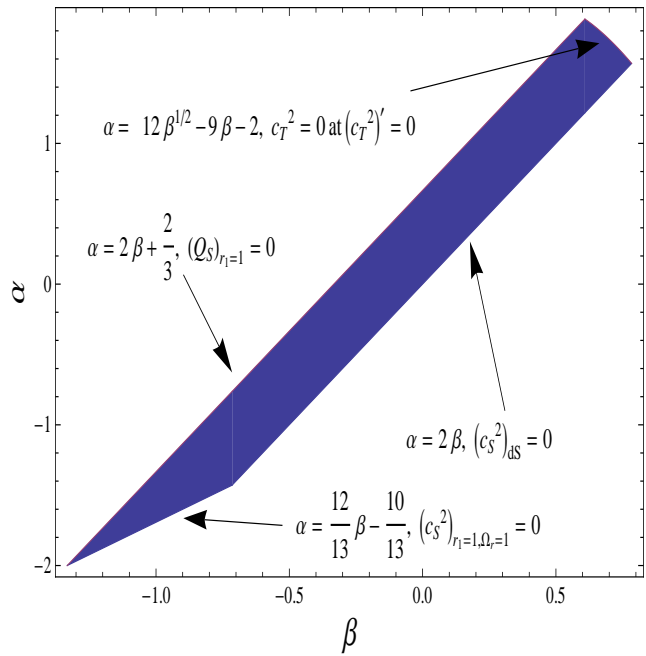} 
\par\end{centering}
\caption{The viable parameter space in the $(\alpha,\beta)$ plane determined
by the conditions (\ref{eq:qcs})-(\ref{eq:qcs4}), (\ref{at3}),
(\ref{at1d}), and (\ref{ctpo}) along the tracker solution $r_{1}=1$
\cite{DT2}.}
\centering{}\label{fig1} 
\end{figure}
%%%%%%%%%%%%%%%%%%%%%%%%%%%%

Let us consider the intermediate regime between $r_{2}\ll1$ and $r_{2}=1$.
As long as the conditions $Q_{S}>0$, $Q_{T}>0$, and $c_{S}^{2}>0$
are satisfied both in the regimes $r_{2}\ll1$ and $r_{2}=1$, the
violation of these conditions does not occur in the intermediate epoch.
However, the tensor propagation speed squared can be negative even
if the conditions (\ref{eq:qcs4}) and (\ref{at4}) are satisfied.
Along the tracker Eq.~(\ref{st4}) gives 
\begin{equation}
c_{T}^{2}=\frac{2+(2-\alpha-9\beta-3\beta\Omega_{r})r_{2}+(9\beta-\alpha)r_{2}^{2}}{(1+r_{2})[2+3(\alpha-2\beta)r_{2}]}\,.\label{cTtr}
\end{equation}
 The transition to the dS solution ($r_{2}=1$) occurs only recently,
so that the term $\Omega_{r}$ can be neglected in Eq.~(\ref{cTtr}).
Then $c_{T}^{2}$ has an extremum at 
\begin{equation}
r_{2}=\frac{4\alpha-15\beta\pm3\sqrt{\beta(30\beta-8\alpha+12\alpha^{2}-15\alpha\beta-18\beta^{2})}}{15\beta-4\alpha+27\alpha\beta-54\beta^{2}}\,.\label{r2so}
\end{equation}
 If $\alpha=1.9$ and $\beta=0.8$, for example, the physical solution
corresponds to the plus sign in Eq.~(\ref{r2so}), i.e. $r_{2}=0.636$,
at which $c_{T}^{2}$ has a minimum. As $\beta$ approaches 1, the
minimum values of $c_{T}^{2}$ get smaller. For $\beta$ around 1,
$c_{T}^{2}$ can be negative for the plus sign of Eq.~(\ref{r2so}).
This leads to the following condition for $\beta>0$: 
\begin{equation}
2\beta<\alpha<12\sqrt{\beta}-9\beta-2\,.\label{ctpo}
\end{equation}
If $\beta<0$, then $c_{T}^{2}$ remains to be positive. Hence we
do not have any additional constraint in the regime $\beta<0$.

In Fig.~\ref{fig1} we plot the parameter space constrained by the
conditions (\ref{eq:qcs})-(\ref{eq:qcs4}), (\ref{at3}), (\ref{at1d}),
and (\ref{ctpo}). For the solutions that start from initial conditions
with $r_{1}\simeq1,r_{2}\ll1$ and then approach the dS attractor
with $r_{1}=1$ and $r_{2}=1$, the parameters $\alpha$ and $\beta$
need to be inside the purple region in Fig.~\ref{fig1}. There is
another case in which both $r_{1}$ and $r_{2}$ are initially much
smaller than 1. We shall address this case in the next subsection.

\subsection{Solutions driven by the term ${\cal L}_{5}$ ($r_{1}\ll1,r_{2}\ll1$)}

{}From Eq.~(\ref{eq:DRr1}), it is clear that another equilibrium
point exists, namely, $r_{1}=0$. Let us now discuss this equilibrium
point in more detail. In this case Eqs.~(\ref{eq:DRr2}) and (\ref{eq:DRr3})
reduce to 
\begin{eqnarray}
r_{2}' & = & -\frac{r_{2}\left(21r_{2}\beta-11\Omega_{r}-3\right)}{21r_{2}\beta+8}\,,\label{eq:r0e1}\\
\Omega_{r}' & = & -\frac{2\Omega_{r}\left(21r_{2}\beta-4\Omega_{r}+4\right)}{21r_{2}\beta+8}\,,\label{eq:r0e2}
\end{eqnarray}
 which depend on $\beta$. The dominant contribution to the field
energy density comes from the term ${\cal L}_{5}$, i.e. $\Omega_{{\rm DE}}=7\beta r_{2}$.

We have the following fixed points 
\begin{equation}
{\rm (A')}~(r_{1},r_{2},\Omega_{r})=(0,0,1)\,,\qquad{\rm (B')}~(r_{1},r_{2},\Omega_{r})=(0,0,0)\,,\qquad{\rm (C')}~(r_{1},r_{2},\Omega_{r})=(0,1/(7\beta),0)\,,\label{eq:fxpr0}
\end{equation}
 which represent radiation, matter, and dark energy dominated points,
respectively. Perturbing Eq.~(\ref{eq:DRr1}) on the $r_{1}=0$ solution
leads to 
\begin{equation}
\delta r_{1}'=\frac{21r_{2}\beta+\Omega_{r}+9}{21r_{2}\beta+8}\,\delta r_{1}\,,\label{eq:unst0}
\end{equation}
 which implies that none of the fixed points (A$'$)-(C$'$) can be
stable. In particular the eigenvalues of the matrix ${\cal M}$, where
$\delta\bm{r}'={\cal M}\,\delta\bm{r}$ and $\delta\bm{r}={}^{t}(\delta r_{1},\delta r_{2},\delta r_{3})$,
are given by 
\begin{equation}
{\rm (A')}~(5/4,7/4,1)\,,\qquad{\rm (B')}~(9/8,3/8,-1)\,,\qquad{\rm (C')}~(12/11,-3/11,-14/11)\,.\label{eq:eigMr0}
\end{equation}
 This shows that the point (A$'$) is unstable, whereas the other
two are saddle. Recalling that the dS fixed point (C) discussed in
the previous subsection is stable against homogenous perturbations,
the solutions finally approach (C) instead of (C$'$). Unless $r_{1}$
is initially very small such that the solutions reach $r_{1}=1$ only
at late times, the system approaches the stable $r_{1}=1$ direction
much before the dS epoch.

In the regime $r_{1}\ll1$ and $r_{2}\ll1$ it is possible to derive
analytic solutions for $r_{1}$ and $r_{2}$ as well as for $w_{{\rm DE}}$
and $w_{{\rm eff}}$. In fact, Eqs.~(\ref{eq:DRr1}), (\ref{eq:DRr2}),
and (\ref{eq:DRr3}) can be simplified as 
\begin{eqnarray}
 &  & r_{1}'\simeq\frac{1}{8}(\Omega_{r}+9)r_{1}\,,\label{r1ap}\\
 &  & r_{2}'\simeq\frac{1}{8}(11\Omega_{r}+3)r_{2}\,,\label{r2ap}\\
 &  & \Omega_{r}'\simeq-\Omega_{r}(1-\Omega_{r})\,,\label{Omeap}
\end{eqnarray}
 where we have assumed that $|\beta|$ is not very much smaller than
unity. During the radiation domination ($\Omega_{r}=1$), integration
of Eqs.~(\ref{r1ap}) and (\ref{r2ap}) gives 
\begin{equation}
r_{1}\propto a^{5/4}\,,\qquad r_{2}\propto a^{7/4}\,,\label{rradiation}
\end{equation}
 whereas during the matter era one has 
\begin{equation}
r_{1}\propto a^{9/8}\,,\qquad r_{2}\propto a^{3/8}\,.\label{rmatter}
\end{equation}
 Eventually the solutions approach the tracker $r_{1}=1$.

In the regime $r_{1}\ll1,r_{2}\ll1$ one has 
\begin{equation}
w_{{\rm DE}}\simeq-(1+\Omega_{r})/8\,,\qquad w_{{\rm eff}}\simeq\Omega_{r}/3\,.\label{wdesm}
\end{equation}
 This gives $w_{{\rm DE}}\simeq-1/4$ and $w_{{\rm eff}}\simeq1/3$
during the radiation era, whereas $w_{{\rm DE}}\simeq-1/8$ and $w_{{\rm eff}}\simeq0$
during the matter era.

The condition (\ref{st1}) reduces to 
\begin{equation}
Q_{S}/M_{{\rm pl}}^{2}\simeq60\beta r_{2}>0\,.\label{Qsre}
\end{equation}
 The sign change of $r_{2}$ implies the appearance of ghosts. For
the initial conditions with $r_{2}>0$ we require that 
\begin{equation}
\beta>0\,.
\end{equation}
 If the solutions start from the regime $r_{1}\ll1,r_{2}\ll1$ and
subsequently enter the regime $r_{1}=1$, the allowed parameter space
in Fig.~\ref{fig1} is restricted be $\beta>0$. Since $Q_{T}/M_{{\rm pl}}^{2}\simeq1/2$,
the no-ghost condition for the tensor mode is automatically satisfied.

The propagation speeds of scalar and tensor perturbations are given,
respectively, by 
\begin{eqnarray}
 &  & c_{S}^{2}\simeq(1+\Omega_{r})/40\,,\label{cssma}\\
 &  & c_{T}^{2}\simeq1+3\beta r_{2}(5-3\Omega_{r})/8\,,\label{ctsma}
\end{eqnarray}
 which are both positive for $0\le\Omega_{r}\le1$. The scalar mode
remains sub-luminal during the radiation era ($c_{S}^{2}=1/20$) and
the matter era ($c_{S}^{2}=1/40$). Under the no-ghost condition (\ref{Qsre})
the tensor mode becomes super-luminal 
(although $c_{T}^{2}$ is very close to 1).

\subsection{Numerical simulations for the cosmological dynamics}

%%%%%%%%%%%%%%%%%%%%%%%%%%%%%
\begin{figure}
\begin{centering}
\includegraphics[width=3.3in,height=3.2in]{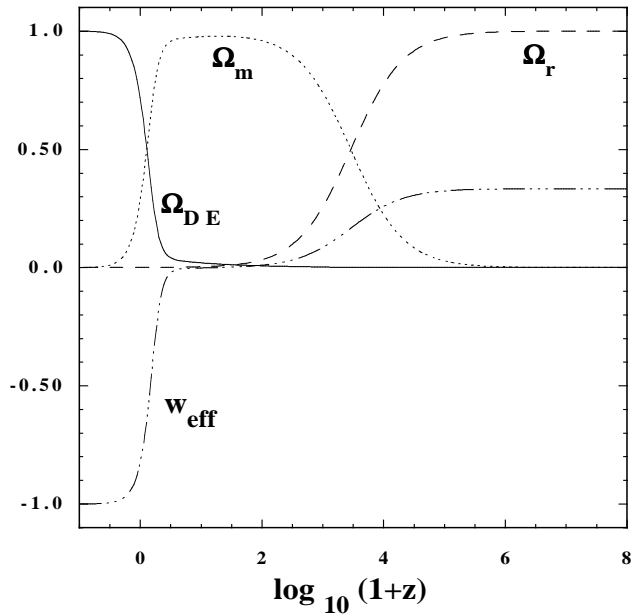} 
\par\end{centering}
\caption{Evolution of $\Omega_{{\rm DE}}$, $\Omega_{m}$, $\Omega_{r}$, and
$w_{{\rm eff}}$ versus the redshift $z=1/a-1$ for $\alpha=0.3$,
$\beta=0.14$, $\epsilon_{2}=1$, $\epsilon_{4}=1$, and $x_{{\rm dS}}=1$.
We choose the initial conditions $r_{1}=1.500\times10^{-10}$, $r_{2}=2.667\times10^{-12}$,
and $\Omega_{r}=0.999992$ at $z=3.63\times10^{8}$.}
\centering{}\label{fig2} 
\end{figure}
%%%%%%%%%%%%%%%%%%%%%%%%%%%%

%%%%%%%%%%%%%%%%%%%%%%%%%%%%%
\begin{figure}
\begin{centering}
\includegraphics[width=3.3in,height=3.2in]{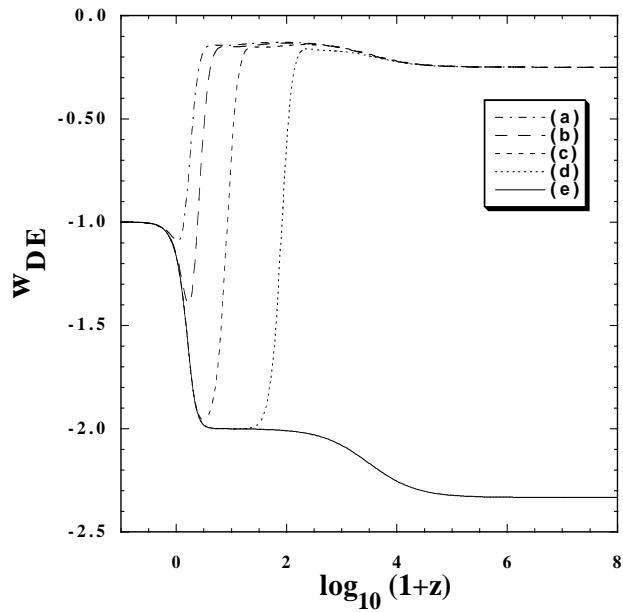} 
\par\end{centering}
\caption{Variation of $w_{{\rm DE}}$ versus $z$ for $\alpha=0.3$, $\beta=0.14$,
$\epsilon_{2}=1$, $\epsilon_{4}=1$, and $x_{{\rm dS}}=1$ {[}cases
(a)-(d){]}. We choose four different initial conditions: (a) $r_{1}=5.000\times10^{-11}$,
$r_{2}=8.000\times10^{-12}$, and $\Omega_{r}=0.999995$ at $z=5.89\times10^{8}$,
(b) $r_{1}=1.500\times10^{-10}$, $r_{2}=2.667\times10^{-12}$, and
$\Omega_{r}=0.999992$ at $z=3.63\times10^{8}$, (c) $r_{1}=5.000\times10^{-9}$,
$r_{2}=8.000\times10^{-14}$, and $\Omega_{r}=0.99995$ at $z=6.72\times10^{7}$,
(d) $r_{1}=5.000\times10^{-6}$, $r_{2}=8.000\times10^{-17}$, and
$\Omega_{r}=0.9986$ at $z=2.04\times10^{6}$. The case (e) corresponds
to $\alpha=-1.5$, $\beta=-0.9$, $\epsilon_{2}=1$, $\epsilon_{4}=-1$,
and $x_{{\rm dS}}=1$ with initial conditions $r_{1}=1$, $r_{2}=10^{-60}$,
and $\Omega_{r}=0.99999$ at $z=3.12\times10^{8}$.}
\centering{}\label{fig3} 
\end{figure}
%%%%%%%%%%%%%%%%%%%%%%%%%%%%

Numerically we integrate Eqs.~(\ref{eq:DRr1})-(\ref{eq:DRr3}) to
confirm the analytic estimation in the previous subsections.

Let us consider the case in which the variables $r_{1}$ and $r_{2}$
are much smaller than 1 at the initial stage of cosmological evolution.
Our numerical simulations show that $r_{1}$ and $r_{2}$ evolve as
Eq.~(\ref{rradiation}) during the radiation era, whereas their evolution
during the matter era is given by Eq.~(\ref{rmatter}). Depending
on the initial conditions of $r_{1}$ and $r_{2}$, the epoch at which
the solutions approach the tracker ($r_{1}=1$) is different. As we
increase the initial ratio $r_{1}/r_{2}$, this epoch tends to occur
earlier. After the solutions reach the tracker, the evolution of $r_{2}$,
$\Omega_{r}$, and $\Omega_{{\rm DE}}$ is given by Eqs.~(\ref{eq:solR2}),
(\ref{eq:omsol}), and (\ref{eq:omede}), respectively.

In Fig.~\ref{fig2} we plot one example for the evolution of density
parameters $\Omega_{{\rm DE}}$, $\Omega_{m}$, and $\Omega_{r}$
as well as the effective equation of state $w_{{\rm eff}}$. In this
case the transition to the regime $r_{1}\simeq1$ occurs only recently,
e.g., $r_{1}=0.99$ around $z=0.07$ with $r_{2}\simeq0.6$. After
passing the present epoch, the solutions are attracted by the dS solution
characterized by $(r_{1},r_{2})=(1,1)$. Figure \ref{fig2} shows
that the sequence of radiation ($\Omega_{r}=1$, $w_{{\rm eff}}=1/3$),
matter ($\Omega_{m}=1$, $w_{{\rm eff}}=0$), and dS ($\Omega_{{\rm DE}}=1$,
$w_{{\rm eff}}=-1$) epochs is in fact realized. Unlike dark energy
models based on $f(R)$ theories, the Galileon model is not plagued
by the presence of a rapidly oscillating mode associated with a heavy
field mass in the early Universe.

Figure \ref{fig3} illustrates the variation of $w_{{\rm DE}}$ for
several different initial conditions and model parameters. The cases
(a)-(d) correspond to $\alpha=0.3$, $\beta=0.14$, $\epsilon_{2}=1$,
$\epsilon_{4}=1$, and $x_{{\rm dS}}=1$ with different initial conditions
satisfying $r_{1}\ll1$ and $r_{2}\ll1$, whereas the case (e) shows
the tracker solution starting from the initial condition $r_{1}=1$
and $r_{2}\ll1$ with the model parameters $\alpha=-1.5$, $\beta=-0.9$,
$\epsilon_{2}=1$, $\epsilon_{4}=-1$, and $x_{{\rm dS}}=1$. Clearly
the solutions with different initial conditions converge to the tracker,
depending on the epoch at which the variable $r_{1}$ grows to the
order of 1. In the cases (a)-(d) the dark energy equation of state
evolves as Eq.~(\ref{wdesm}) in the regime $r_{1}\ll1$ and $r_{2}\ll1$
($w_{{\rm DE}}\simeq-1/4$ and $w_{{\rm DE}}\simeq-1/8$ during the
radiation and matter eras, respectively), which is followed by the
evolution given in Eq.~(\ref{wdetra}) after the solutions reach
the tracker at $r_{1}=1$. As long as the tracking behavior occurs
by today, the dark energy equation of state crosses the cosmological
constant boundary ($w_{{\rm DE}}=-1$).

Numerically we find that for the initial conditions with $r_{1}\lesssim2$
the solutions are typically attracted by the tracker. On the other
hand, if $r_{1}\gtrsim2$, the system tends to approach the matter-dominated
epoch with the growth of $r_{1}$. In the latter case the dominant
contribution to $\Omega_{{\rm DE}}$ comes from the term ${\cal L}_{2}$,
so that $\Omega_{{\rm DE}}$ decreases as in quintessence without
a potential.

%%%%%%%%%%%%%%%%%%%%%%%%%%%%%
\begin{figure}
\begin{centering}
\includegraphics[width=3.3in,height=3.1in]{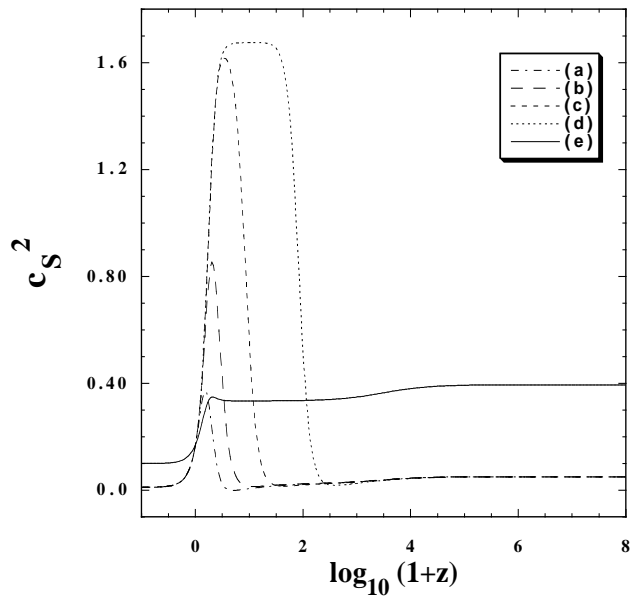} 
\par\end{centering}
\caption{Evolution of $c_{S}^{2}$ versus $z$ for the same model parameters
and initial conditions as given in Fig.~\ref{fig3}.}
\centering{}\label{fig4} 
\end{figure}
%%%%%%%%%%%%%%%%%%%%%%%%%%%%

In Fig.~\ref{fig4} we plot the evolution of $c_{S}^{2}$ for the
same model parameters and initial conditions as those presented in
Fig.~\ref{fig3}. In the regime $r_{1}\ll1$ and $r_{2}\ll1$, our
numerical simulations in the cases (a)-(d) agree with the analytic
estimation of the scalar propagation speed given in Eq.~(\ref{cssma}),
i.e. $c_{S}^{2}\simeq1/20$ and $c_{S}^{2}\simeq1/40$ during the
radiation and matter eras respectively. As the solutions reach the
regime $r_{1}\simeq1$ with $r_{2}\ll1$, $c_{S}^{2}$ approaches
the value estimated by Eq.~(\ref{at3}). When $\alpha=0.3$ and $\beta=0.14$
the analytic estimation gives $c_{S}^{2}\simeq1.67$ during the matter
dominance, which agrees with the value at the plateau in the case
(d) of Fig.~\ref{fig4}. Finally the solutions reach the dS fixed
point, at which $c_{S}^{2}$ shifts to the value given in Eq.~(\ref{csds}),
e.g., $c_{S}^{2}=1.01\times10^{-2}$ for $\alpha=0.3$ and $\beta=0.14$.

For positive $\beta$ one can show that under the conditions (\ref{eq:qcs}),
(\ref{csds}), and (\ref{at1}) the scalar propagation speed estimated
by Eq.~(\ref{at3}) becomes super-luminal. However, the scalar mode
can remain sub-luminal provided the solutions reach the regime $r_{1}=1$
in the recent past. The cases (a) and (b) in Fig.~\ref{fig4} correspond
to such examples in which the peak value of $c_{S}^{2}$ is smaller
than 1.

If $\beta<0$ there is a parameter space in which the scalar propagation
speed (\ref{at3}) is sub-luminal, while satisfying the conditions
(\ref{eq:qcs}), (\ref{csds}), and (\ref{at1}). In this case the
initial conditions of $r_{1}$ need to be close to 1. If $r_{1}$
is smaller than the order of unity, the scalar ghost appears for negative
$\beta$. On the other hand, if $r_{1}\gtrsim2$, the solutions do
not finally approach the dS fixed point. The case (e) in Fig.~\ref{fig4}
corresponds to an example of the sub-luminal evolution of $c_{S}^{2}$
for negative $\beta$ with the initial condition $r_{1}=1$. Since
the solution stays on the tracker, the scalar propagation speed is
given by Eq.~(\ref{at3}) during the radiation and matter eras and
by Eq.~(\ref{csds}) at the dS point.

%%%%%%%%%%%%%%%%%%%%%%%%%%%%%
\begin{figure}
\begin{centering}
\includegraphics[width=3.3in,height=3.2in]{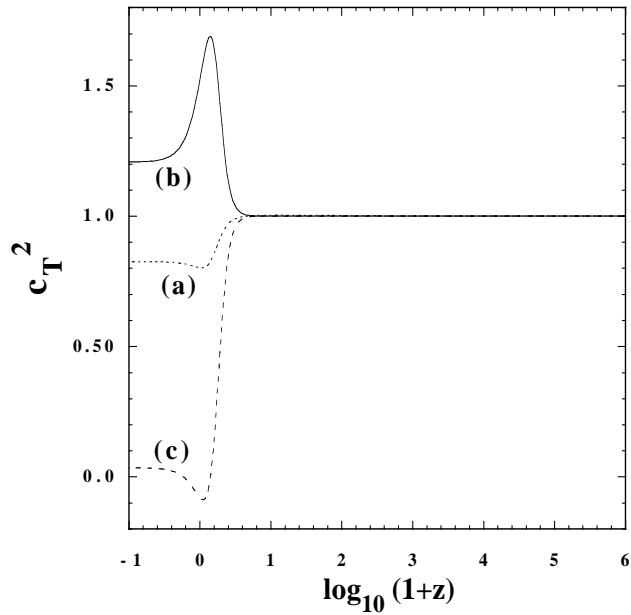} 
\par\end{centering}
\caption{Evolution of $c_{T}^{2}$ versus $z$ for two cases: (a) $\alpha=0.3$,
$\beta=0.14$, $\epsilon_{2}=1$, $\epsilon_{4}=1$, and $x_{{\rm dS}}=1$
with initial conditions $r_{1}=1.500\times10^{-10}$, $r_{2}=2.667\times10^{-12}$,
and $\Omega_{r}=0.999992$ at $z=3.63\times10^{8}$, (b) $\alpha=-1.5$,
$\beta=-0.9$, $\epsilon_{2}=1$, $\epsilon_{4}=-1$, and $x_{{\rm dS}}=1$
with initial conditions $r_{1}=1$, $r_{2}=10^{-60}$, and $\Omega_{r}=0.99999$
at $z=3.12\times10^{8}$, and (c) $\alpha=1.9$, $\beta=0.8$, $\epsilon_{2}=1$,
$\epsilon_{4}=1$, and $x_{{\rm dS}}=1$ with initial conditions $r_{1}=10^{-5}$,
$r_{2}=10^{-35}$, and $\Omega_{r}=0.99999$ at $z=3.12\times10^{8}$.}
\centering{}\label{fig5} 
\end{figure}
%%%%%%%%%%%%%%%%%%%%%%%%%%%%

For the initial conditions with $r_{1}\ll1$ and $r_{2}\ll1$ the
tensor propagation speed starts to evolve from the value estimated
by Eq.~(\ref{ctsma}), which is slightly super-luminal under the
no-ghost condition $\beta>0$ for the scalar mode. After the solutions
reach the regime $r_{1}\simeq1$ and $r_{2}\ll1$, $c_{T}^{2}$ is
still close to 1 because it is described by Eq.~(\ref{at4}). The
tensor propagation speed finally approaches the value (\ref{eq:qcs4})
at the dS point. During the transition from the regime $r_{2}\ll1$
to the regime $r_{2}\simeq1$, there is an epoch at which $c_{T}^{2}$
can have either the maximum or the minimum. In the case (a) of Fig.~\ref{fig5},
the analytic formulas in Eqs.~(\ref{cTtr}) and (\ref{r2so}) show
that $c_{T}^{2}$ has a minimum value $0.799$ at $r_{2}=0.662$ {[}plus
sign of Eq.~(\ref{r2so}){]}, whereas in the case (b) $c_{T}^{2}$
possesses a maximum value $1.690$ at $r_{2}=0.412$ {[}minus sign
of Eq.~(\ref{r2so}){]}. This estimation agrees well with the numerical
results shown in Fig.~\ref{fig5}. In the case (c) of Fig.~\ref{fig5}
the condition (\ref{ctpo}) is violated, so that $c_{T}^{2}$ has
a negative minimum. In the region where $\alpha$ and $\beta$ are
positive, the condition (\ref{ctpo}) needs to be satisfied to avoid
the temporal Laplacian instability of the tensor mode.

If the solutions start from the regime $r_{1}\simeq1$ and $r_{2}\ll1$,
then the tensor propagation speed (\ref{at4}) can be sub-luminal
under the condition $4\alpha+3\beta+3\beta\Omega_{r}>0$ for the branch
$r_{2}>0$. In this case, however, $c_{T}^{2}$ exceeds 1 at the dS
point, as long as the conditions (\ref{eq:qcs})-(\ref{eq:qcs3})
and (\ref{at1})-(\ref{at3}) are satisfied. Since $c_{T}^{2}>1$
in the regime $r_{1}\ll1$ and $r_{2}\ll1$ as well, it is not possible
to avoid the appearance of the super-luminal mode for tensor perturbations.
However, the super-luminal propagation does not necessarily imply
the inconsistency of Galileon theory because of the possibility for
the absence of the closed causal curve \cite{Sami}.

%%%%%%%%%%%%%%%%%%%%%%%%%%%%%%%%%%%%%%%%%%%%%%%%%%%%%%%%%%%%%%%%%%%%%%%%%%%%%
\section{Cosmology based on the models with non-constant functions $F(\phi)$}
%%%%%%%%%%%%%%%%%%%%%%%%%%%%%%%%%%%%%%%%%%%%%%%%%%%%%%%%%%%%%%%%%%%%%%%%%%%%%

We shall proceed to the cosmology for the theories with non-constant 
$F$ in which the functions $F$ and $f_i$ ($i=1, 2, \cdots, 5$) are given 
in Eq.~(\ref{funchoice2}) with $d_{\xi}=0$.
We take into account radiation ($\rho_1=\rho_r$, $w_1=1/3$) and 
non-relativistic matter ($\rho_2=\rho_m$, $w_2=0$), which 
satisfy the continuity equations (\ref{be3}).
Taking the time-derivative of Eq.~(\ref{eq:frd1}) and combining it 
with Eq.~(\ref{eq:frd2}), we obtain the equations of motion for
$\ddot{\phi}$ and $\dot{H}$.
Then the dimensionless variables $\tilde{x}, \tilde{y}$
defined in Eq.~(\ref{tildexy}) and the radiation density 
parameter $\Omega_r=\rho_r/(3F H^2)$ obey the following equations
\begin{eqnarray}
\tx' &=& \tx [ 29160000000000\,
\tx^{3} 
( -8\,\Omega_r\, ( 12+d_2^2 \epsilon_2 \,p \tx ) 
+ ( 4+ ( -4+d_2^{2} \epsilon_2 ) p \tx )  
( 24+px ( 24+d_2^2 \epsilon_2 \,p \tx ))) 
\nonumber \\
& &-1440000000000\,d_3^{3}{p}^{2} \tx^{3} 
( -216-216\,\Omega_r+p \tilde{x} ( 24+p \tx 
( -192+d_2^{2} \epsilon_2\, ( 45+p \tx )))) 
\ty^2
\nonumber \\
& &-625000000\,{p}^{3}\tx^{2} 
(2048\,d_3^{6}{p}^{2}\tx^{2} ( -9+p \tx ) 
+6561\,d_4^{4}\epsilon_4\, ( 144+80\,\Omega_r
+p \tx ( 20-p\tx ( -76+d_2^{2} \epsilon_2
\, ( 19+p \tx ))))) \ty^{4} 
\nonumber \\
& &+583200000\,{p}^{4} \tx ( 3125\,d_3^{3}d_4^{4}
\epsilon_4\,{p}^{2}\tx^{2} ( -11+p \tx ) 
+192\,d_5^{5} ( 600+280\,\Omega_r+p \tx 
( 88+p \tx ( 272 \nonumber \\
& &-d_2^{2}\epsilon_2\,
( 73+5\,p \tx ))))) \ty^{6}
-84375\,{p}^{7} \tx^{2} ( 11390625\,d_4^{8}\epsilon_4^{2}
( -6+p \tx ) +524288\,d_3^{3}d_5^{5} ( -45
+2\,p \tx )) \ty^{8}
\nonumber \\
& &+15746400000\,d_4^{4}d_5^
{5}\epsilon_4\,{p}^{8} \tx ( -59+9\,p \tx ) \ty^{10}
-6019743744\,d_5^{10}{p}^{9} (-5+p \tx ) 
\ty^{12}]/(60\Delta)\,, 
\label{xgene} \\
\ty' &=& \tx \ty [ 116640000000000 \tx^2 (24-24 \Omega_r
+p \tx (-48-24 p\tx+d_2^2 \epsilon_2 (12+5p \tx))
-4320000000000 d_3^3 p^2 \tx^2 (72-24\Omega_r \nonumber \\
& &+p \tx (-16+(-32+3d_2^2 \epsilon_2) p\tx )) \ty^2
-625000000 p^3 \tx (2048 d_3^6 p^3 \tx^3+2187 d_4^4 
\epsilon_4 (96 \Omega_r+p \tx (108-p \tx (-120 \nonumber \\
& &+d_2^2 \epsilon_2 (18+p \tx))))) \ty^4
+64800000 p^4 (3125 d_3^3 d_4^4 \epsilon_4 p^2 \tx^2 (-9+10p \tx)
+1728 d_5^5 (120+120 \Omega_r \nonumber \\
& &+p \tx (128+p \tx (144
-d_2^2 \epsilon_2 (27+2p \tx))))) \ty^6
-253125 p^7 \tx (3796875 d_4^8 \epsilon_4^2 p \tx
+524288 d_3^3 d_5^5 (-3+p \tx)) \ty^8
\nonumber \\
& &+5248800000 d_4^4 d_5^5 \epsilon_4 p^8 (-9+28p \tx)
\ty^{10}-6019743744 d_5^{10} p^{10} \ty^{12}]/(60 \Delta)\,,
\\
\Omega_r' &=&
p\,\Omega_r [972000000000 \tx^4 (-48p \tx+d_2^2 \epsilon_2 
(8-8\Omega_r +p \tx (16+(-4+d_2^2 \epsilon_2) p \tx))) 
-48000000000 d_3^3 p \tx^3 (144-144 \Omega_r
\nonumber \\
& &+p\tx (-72+p \tx (-168+d_2^2 \epsilon_2 (36+p\tx)))) \ty^2
+62500000 p^2 \tx^2 (-1024 d_3^6 p^2 \tx^2 (-2+p \tx)+729d_4^4
\epsilon_4 (144-144 \Omega_r
\nonumber \\
& &+p \tx (-192+p \tx (-204+d_2^2 \epsilon_2 (43+3p
\tx))))) \ty^4+2160000 p^3 \tx (3125 d_3^3 d_4^4 \epsilon_4 
p^2 \tx^2 (-18+17 p \tx)+1728 d_5^5 (-160 \nonumber \\
& & +160 \Omega_r+p \tx 
(280+p \tx (248-d_2^2 \epsilon_2 (54+5p \tx)))))\ty^6-5625 
p^6 \tx^2 (11390625 d_4^8 \epsilon_4^2 (1+p \tx) \nonumber \\
& &+65536 d_3^3 d_5^5 (-36+23p \tx)) \ty^8
+174960000 d_4^4 d_5^5 \epsilon_4 p^7 \tx 
(88+63p \tx) \ty^{10}-501645312 d_5^{10} p^8 
(2+p \tx) \ty^{12}]/\Delta\,,
\label{Omergene}
\end{eqnarray}
where
\begin{eqnarray}
\Delta &\equiv& p[501645312 d_5^{10}p^8 \ty^{12}-3732480000
d_5^5 p^3 \tx \ty^6 (-160+3d_4^4 \epsilon_4 p^4 \ty^4)
-62500000 \tx^4 (-1024 (-27+d_3^3 p^2 \ty^2)^2 \nonumber \\
& &+243 d_2^2 \epsilon_2 (512+3d_4^4 \epsilon_4 p^4
\ty^4 ))+151875 p^2 \tx^2 \ty^4 (65536 d _5^5 p^2 \ty^2
(-45+d_3^3 p^2 \ty^2)+84375 d_4^4 \epsilon_4 (-512+
5 d_4^4 \epsilon_4 p^4 \ty^4)) \nonumber \\
& &+4320000 p \tx^3 \ty^2 (108p^2 \ty^2 (9375 d_4^4 \epsilon_4
+16 d_2^2 d_5^5 \epsilon_2 p^2 \ty^2)
-3125 d_3^3 (-512+9 d_4^4 \epsilon_4 p^4 \ty^4))]\,.
\end{eqnarray}

The dS fixed point with $\tx=\tx_{\rm dS}=\ty$ 
and $\Omega_r=0$ exists under the conditions 
(\ref{gened2}) and (\ref{gened3}).
Since the theory has a nonminimal coupling $F(\phi)R$, 
it is possible to place constraints on the values of $x$
around today from the variation of the effective gravitational 
coupling, $G_{\rm eff}=[8\pi F(\phi)]^{-1}$. 
The Lunar Laser Ranging experiments give the bound
$|\dot{G}_{\rm eff}/G_{\rm eff}|<1.3 \times 10^{-12}$\,yr$^{-1}$ \cite{Lunar}, 
or in terms of the present Hubble parameter $H_0$, 
$|\dot{G}_{\rm eff}/G_{\rm eff}|<0.02H_0$ \cite{Babi11}.
In our theory $|\dot{G}_{\rm eff}/G_{\rm eff}|=|p \tx|\,H$, which
gives the constraint $|p \tx|<0.02$ around today.
Since the value of $\tx_{\rm dS}$ is not much different from 
$\tx$ today, we employ the following criterion
\begin{equation}
|p \tx_{\rm dS}|<{\cal O}(0.01)\,.
\label{genecon1}
\end{equation}
Under this bound the condition (\ref{dsstability}) is always satisfied, 
which means that the dS solution is classically stable.
From Eq.~(\ref{eq:Cc2sx}) the Laplacian instability of scalar perturbations
at the dS point can be avoided for 
$p \tilde{x}_{\rm dS}<0$.
The no-ghost condition (\ref{eq:Qc2sx}) is satisfied 
provided that $(\phi/M_{\rm pl})^p>0$.

\subsection{Initial conditions with $\ty^2 \gg |\tx|$}

If $\ty^2 \gg |\tx|$ in the early cosmological epoch, 
then the term ${\cal L}_5$ dominates over 
the terms ${\cal L}_{2,3,4}$,
i.e. $\Omega_{\rm DE} \simeq -p \tx 
+(7/3125)d_5^5 p^5 \ty^6/\tx$.
In order to avoid the dominance of dark energy during the 
radiation and matter eras we require that $|p \tx| \ll 1$
and $|d_5^5 p^5 \ty^6| \ll |\tx|$.
In this regime the quantities $Q_S$ and $Q_T$ 
defined in Eqs.~(\ref{eq:genQs}) and (\ref{eq:genQW})
are approximately given by 
\begin{eqnarray}
& & \frac{Q_S}{M_{\rm pl}^2} \simeq \frac{12}{625}
\frac{d_5^5 p^5 \ty^6}{\tx} \left( \frac{\phi}
{M_{\rm pl}} \right)^p\,,\qquad
\frac{Q_T}{M_{\rm pl}^2} \simeq \frac12 \left( 
\frac{\phi}{M_{\rm pl}} \right)^p
\left( 1- \frac{3}{3125}\frac{d_5^5 p^5 \ty^6}
{\tx} \right)\,.
\label{Qsasy}
\end{eqnarray}
The tensor ghost is absent 
for $(\phi/M_{\rm pl})^p>0$.
Since the evolution of the field is given by 
$\phi=\phi_i \exp(\int_{N_i}^N \tx 
d \tilde{N})$, where $\phi_i$ is the initial field value at $N=N_i$, 
the condition $(\phi/M_{\rm pl})^p>0$ is satisfied
for $\phi_i>0$. For the avoidance of the scalar ghost 
we require that 
\begin{equation}
d_5 p \tx>0\,.
\label{genecon3}
\end{equation}
In the regime $\ty^2 \gg |\tx|$ and $|d_5^5 p^5 \ty^6| \ll |\tx|$ 
the scalar and tensor propagation speeds  
defined in Eqs.~(\ref{cS2}) and (\ref{eq:noinstGW})
can be estimated as
\begin{equation}
c_S^2 \simeq \frac{1}{40} (1+\Omega_r)
+\frac{375}{8}(1-\Omega_r) 
\frac{\tx^2}{d_5^5 p^4 \ty^6}\,,
\qquad
c_T^2 \simeq 1+\frac{3}{25000}(4-3\Omega_r)
\frac{d_5^5 p^5 \ty^6}{\tx}\,.
\label{csctana}
\end{equation}
Since $c_T^2$ is close to 1, the tensor instability 
can be avoided. If $d_5>0$, then there is no
instability for the scalar perturbation ($c_S^2>0$).
In the regime $\tx^2 \ll |d_5^5 p^4 \ty^6|$ we have 
$c_S^2 \simeq (1+\Omega_r)/40>0$. 
If $\tx^2 \gtrsim |d_5^5 p^4 \ty^6|$,
it can happen that the scalar perturbation is 
subject to the Laplacian instability for 
negative $d_5$. 

%%%%%%%%%%%%%%%%%%%%%%%%%%%%%%%%%%%%%%
\begin{figure}
\begin{centering}
\includegraphics[width=3.3in,height=3.2in]{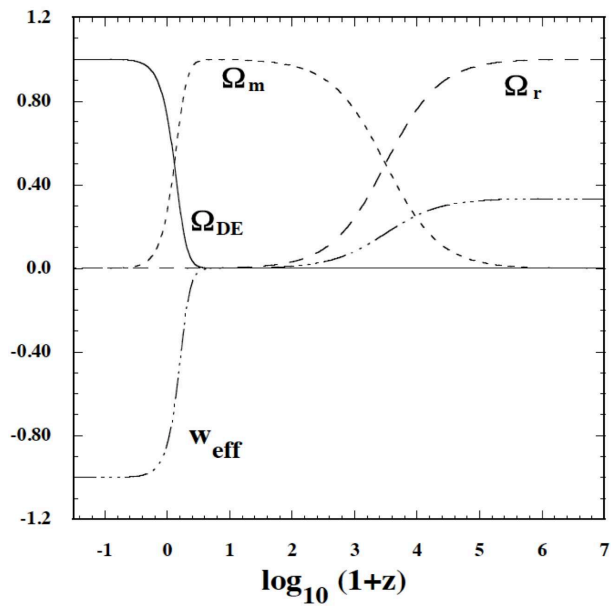} 
\par\end{centering}
\caption{
Evolution of $\Omega_{{\rm DE}}$, $\Omega_{m}$, $\Omega_{r}$, and
$w_{{\rm eff}}$ versus the redshift $z$ for the model with $p=1$, 
$\epsilon_2=1$, $\epsilon_4=-1$, $d_4=1$, $d_5=1$, $d_{\xi}=0$, 
and $\tilde{x}_{\rm dS}=0.007$.
The initial conditions are chosen to be 
$\tilde{x}=1.0 \times 10^{-18}$, $\tilde{y}=1.5 \times 10^{-5}$,
and $\Omega_{r}=0.99992$ at $z=3.9 \times 10^{7}$.}
\centering{}\label{fig6} 
\end{figure}
%%%%%%%%%%%%%%%%%%%%%%%%%%%%%%%%%%%%%%

%%%%%%%%%%%%%%%%%%%%%%%%%%%%%%%%%%%%%%
\begin{figure}
\begin{centering}
\includegraphics[width=3.3in,height=3.2in]{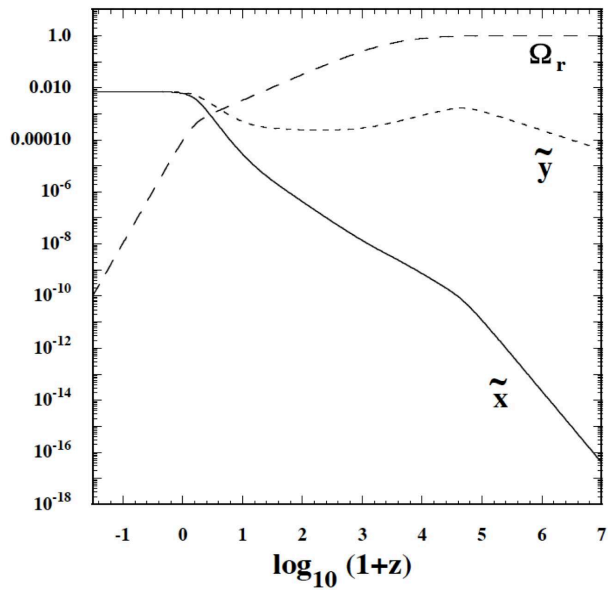} 
\par\end{centering}
\caption{
Variation of $\tilde{x}$, $\tilde{y}$, and $\Omega_r$
versus $z$ for the same model parameters and the initial 
conditions as those given in Fig.~\ref{fig6}.
The solution finally approaches the de Sitter attractor 
with $\tilde{x}=\tilde{x}_{\rm dS}=0.007$.}
\centering{}\label{fig7} 
\end{figure}
%%%%%%%%%%%%%%%%%%%%%%%%%%%%%%%%%%%%%%

In the regime $\ty^2 \gg |\tx|$ and $|d_5^5 p^5 \ty^6| \ll |\tx|$
the autonomous equations (\ref{xgene})-(\ref{Omergene}) are simplified as 
\begin{eqnarray}
\tx' &\simeq& \frac{1}{8}\tx
\left[15+7\Omega_r+625(1-\Omega_r)\frac{\tx^2}{d_5^5 p^4 \ty^6} \right]\,,
\label{txap1}\\
\ty' &\simeq& \frac{3}{8}\ty \left[ 1+\Omega_r+
\frac{625}{3} (1-\Omega_r)\frac{\tx^2}{d_5^5 p^4 \ty^6} \right]\,,
\label{tyap1} \\
\Omega_r' &\simeq& -\Omega_r (1-\Omega_r)\,.
\label{Omerap}
\end{eqnarray}
{}From Eq.~(\ref{Omerap}) there are two fixed points characterized by  
$\Omega_r=1$ and $\Omega_r=0$.
As long as the condition $\tx^2  \ll |d_5^5 p^4 \ty^6|$ is satisfied, 
the evolution of the variables $\tx$ and $\ty$ during the 
radiation era ($\Omega_r=1$) is given by 
\begin{equation}
\tx \propto a^{11/4}\,,\qquad
\ty \propto a^{3/4}\,,
\label{xandyasy}
\end{equation}
whereas during the matter era ($\Omega_r=0$) one has 
\begin{equation}
\tx \propto a^{15/8}\,,\qquad
\ty \propto a^{3/8}\,.
\label{xandyasy2}
\end{equation}
In both cases $\tx$ grows faster than $\ty$.
If the quantity $\tx^2/(d_5^5 p^4 \ty^6)$
becomes larger than the order of unity,
the evolution of $\tx$ and $\ty$ is subject to change.

%%%%%%%%%%%%%%%%%%%%%%%%%%%%%%%%%%%%%%
\begin{figure}
\begin{centering}
\includegraphics[width=3.3in,height=3.2in]{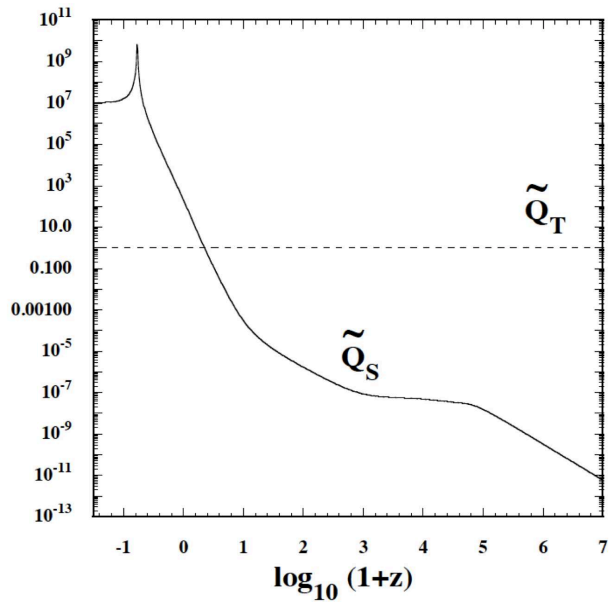} 
\par\end{centering}
\caption{
Evolution of the dimensionless variables 
$\tilde{Q}_S \equiv 2Q_S/(M_{\rm pl}^{2-p}\phi^p)$ 
and $\tilde{Q}_T \equiv 2Q_T/(M_{\rm pl}^{2-p}\phi^p)$
versus $z$ for the same model parameters and 
the initial conditions as those given in Fig.~\ref{fig6}.
The signs of $Q_S$ and $Q_T$ remain to be positive.}
\centering{}\label{fig8} 
\end{figure}
%%%%%%%%%%%%%%%%%%%%%%%%%%%%%%%%%%%%%%

%%%%%%%%%%%%%%%%%%%%%%%%%%%%%%%%%%%%
\begin{figure}
\begin{centering}
\includegraphics[width=3.2in,height=3.2in]{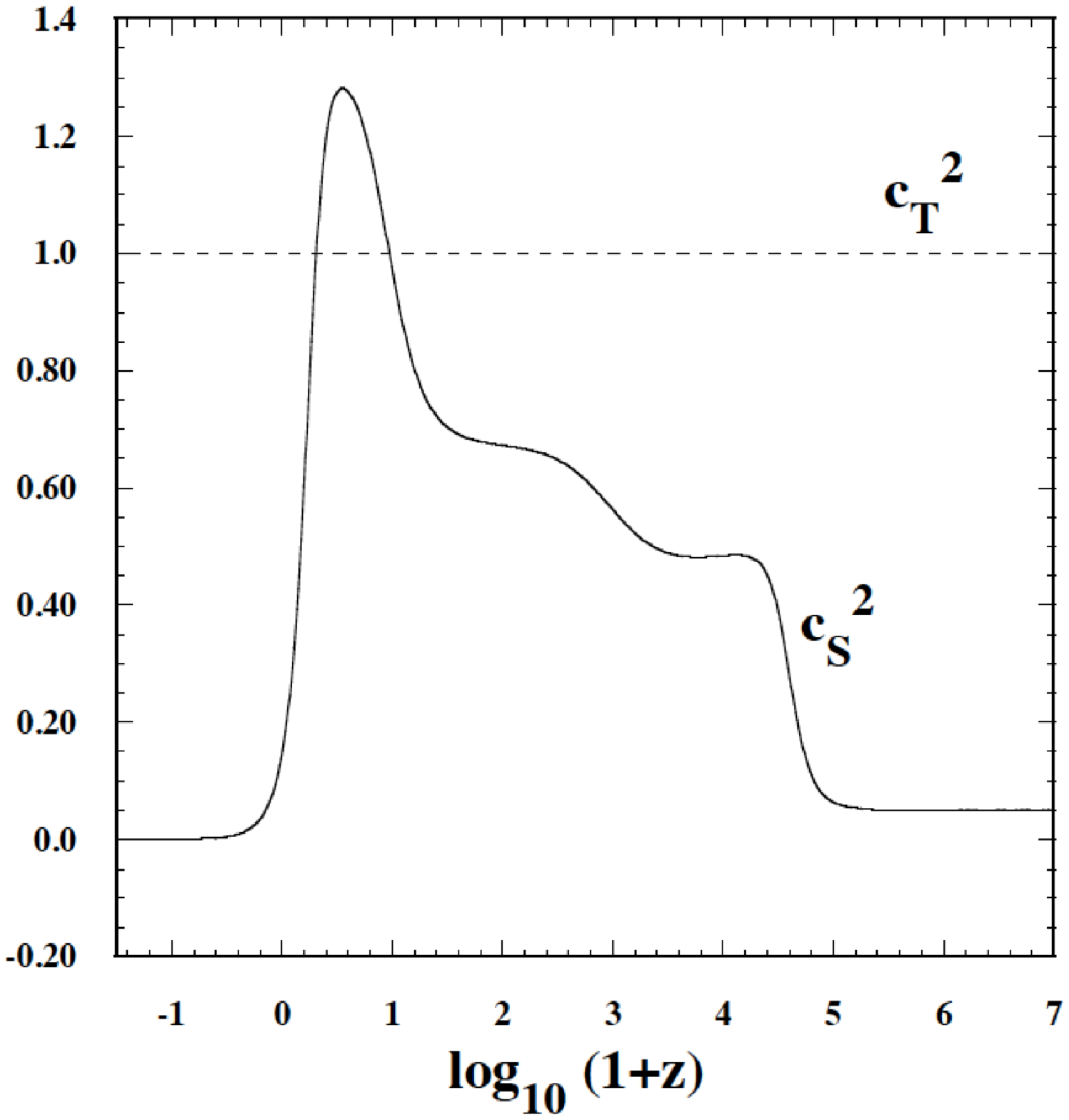} 
\includegraphics[width=3.2in,height=3.2in]{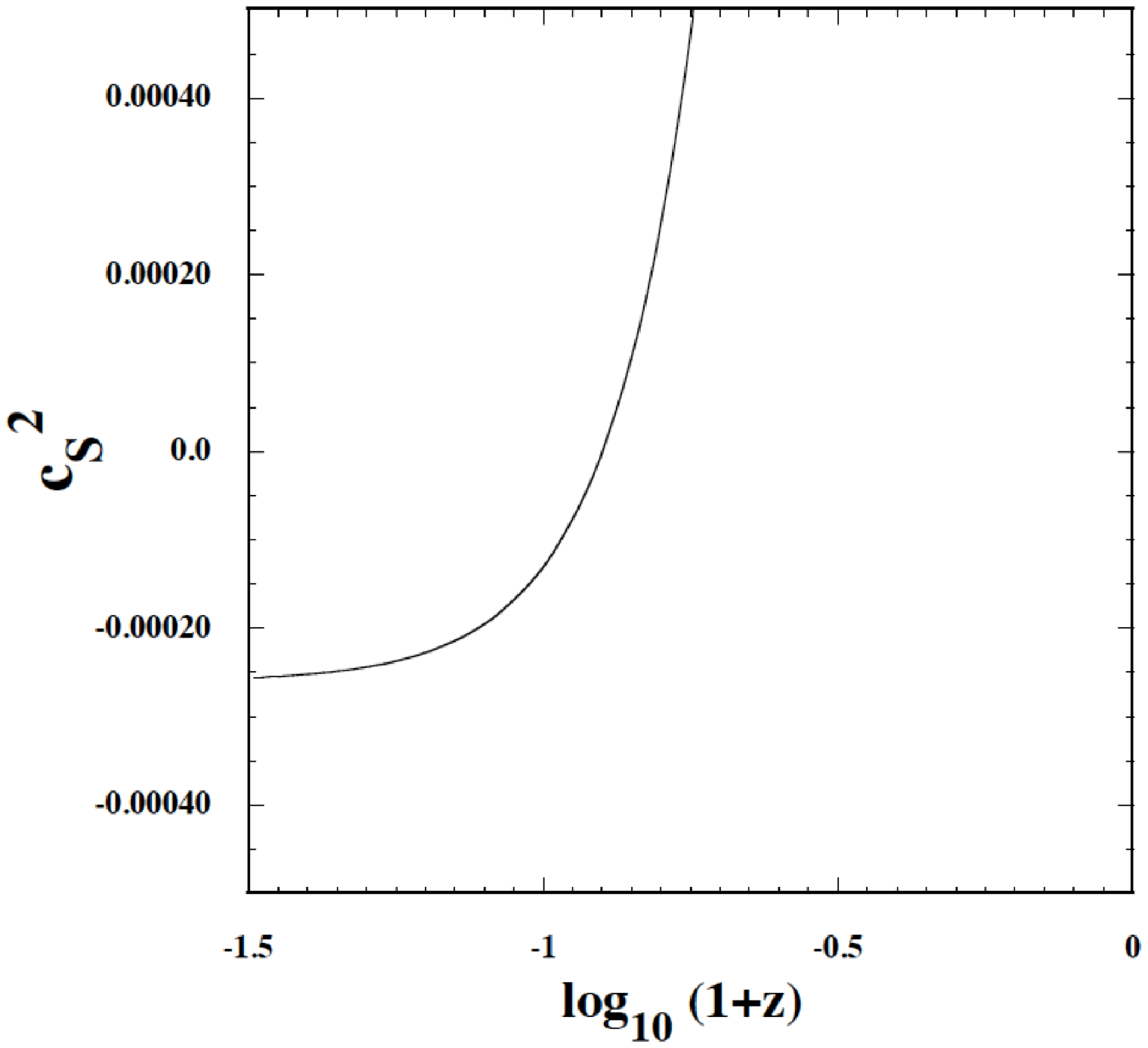} 
\par\end{centering}
\caption{(Left) Evolution of $c_{S}^{2}$ and $c_T^2$ 
versus $z$ for the same model parameters and the initial 
conditions as those given in Fig.~\ref{fig6}.
(Right) The enlarged version for the evolution of 
$c_S^2$ in the regime $-1.5<\log_{10} (1+z)<0$. 
The field propagation speed becomes negative 
in the future (around $z\approx-0.87$).}
\centering{}\label{fig9} 
\end{figure}
%%%%%%%%%%%%%%%%%%%%%%%%%%%%%%%%%%%%

For the solutions starting from the regime $|\ty^2| \gg |\tx|$
the condition (\ref{genecon3}) needs to be satisfied initially.
Then there are two possible cases: 
(i) $p \tx>0$ and $d_5>0$, and (ii) $p \tx<0$ and $d_5<0$.
The avoidance of the scalar Laplacian instability
at the future dS fixed point requires that $p \tx_{\rm dS}<0$.
However, if we demand the viable cosmology {\it by today} 
(the redshift $z \ge 0$), 
the condition $p \tx_{\rm dS}<0$ is not necessarily mandatory.
In general, if the variable $\tx$ changes its sign during the cosmic
expansion history, this signals the violation of the conditions
for no ghosts and no Laplacian instabilities.
For example, this can be seen in the expression of 
$Q_S$ and $c_S^2$ in Eqs.~(\ref{Qsasy}) and (\ref{csctana})
in the past asymptotic regime.
In fact we have numerically confirmed the violation of 
at least one of those conditions.
In the following we shall study the cosmological dynamics
in which the sign of $p \tx$ at the early epoch is same as that 
of $p \tx_{\rm dS}$.
In the case (i) the condition $p \tx_{\rm dS}<0$ is violated, 
but it is possible to realize cosmological trajectories
in which all the required conditions are satisfied by today.
In the case (ii) the condition $p \tx_{\rm dS}<0$ is met, 
but we need to check whether there are no 
violations of the no-ghost and stability conditions
in the cosmic expansion history.

Let us first discuss the cosmological dynamics in the case (i)
with $p \tx_{\rm dS}>0$.
In Fig.~\ref{fig6} we plot the variation of 
$\Omega_{{\rm DE}}$, $\Omega_{m}$, $\Omega_{r}$, and
$w_{{\rm eff}}$ for the model with $p=1$, 
$\epsilon_2=1$, $\epsilon_4=-1$, $d_4=1$, $d_5=1$, $d_{\xi}=0$, 
and $\tilde{x}_{\rm dS}=0.007$ (in which case the condition 
(\ref{genecon1}) is satisfied).
The constants $d_2$ and $d_3$ are known from 
Eqs.~(\ref{gened2}) and (\ref{gened3}).
We choose the initial conditions
$\tx=1.0 \times 10^{-18}$, $\ty=1.5 \times 10^{-5}$,
and $\Omega_{r}=0.99992$ at the redshift $z=3.9 \times 10^{7}$, 
in which case $|\ty^2| \gg |\tx|$ and 
$|d_5^5 p^5 \ty^6| \ll |\tx|$ initially.
The background evolution in Fig.~\ref{fig6} shows that 
the sequence of radiation, matter, and dS eras is
realized in this case.

Figure \ref{fig7} illustrates the evolution of the variables 
$\tx$ and $\ty$ as well as $\Omega_r$.
We find that $\tx$ approaches the dS attractor with $\tx_{\rm dS}=0.007$
without changing its sign. In the regime 
$\tx^2 \ll |d_5^5 p^4 \ty^6|$ the evolution of 
$\tx$ and $\ty$ is well described by the analytic 
estimation (\ref{xandyasy}) during the radiation era.
However, around $z \lesssim 10^5$, the last terms 
in Eqs.~(\ref{txap1}) and (\ref{tyap1}) starts to give 
rise to the contribution to the evolution of $\tx$ and $\ty$. 
As we see in Fig.~\ref{fig7}, $\tx$ and $\ty$ evolve differently 
from the analytic estimation (\ref{xandyasy}) and (\ref{xandyasy2})
for $z \lesssim 10^5$.

In Figs.~\ref{fig8} and \ref{fig9} we plot the variation of the quantities
$\tilde{Q}_S=2Q_S/(M_{\rm pl}^{2-p}\phi^p)$, 
$\tilde{Q}_T=2Q_T/(M_{\rm pl}^{2-p}\phi^p)$, 
$c_S^2$, and $c_T^2$ for the same model parameters and 
initial conditions as those 
given in Fig.~\ref{fig6}.
We find that $\tilde{Q}_S$ grows rapidly, whereas $\tilde{Q}_T$
is always close to 1. Since both $\tilde{Q}_S$ and $\tilde{Q}_T$
are positive, the appearance of the scalar and tensor ghosts
is avoided in this case.

Figure \ref{fig9} shows that $c_S^2$ starts to evolve from the 
value around $0.05$, as estimated analytically in 
Eq.~(\ref{csctana}). For $z \lesssim 10^5$ the 
contribution of the second term in the expression of $c_S^2$
in Eq.~(\ref{csctana}) becomes important, which leads 
to the increase of $c_S^2$.
For the model parameters given in Fig.~\ref{fig6}
the scalar propagation speed slightly exceeds 1 
during the transition from the matter era to the
dS epoch. In Fig.~\ref{fig9} we find that 
$c_S^2$ remains positive until recently ($z \ge 0$).
However, since the sign of $\tx$ is always positive, 
$c_S^2$ is negative at the dS point, i.e.
$c_S^2 \simeq -p\tx_{\rm dS}/27=-2.6 \times 10^{-4}$. 
The crossing of $c_S^2$ at 0 occurs in future
around the redshift $z\approx-0.87$.
The tensor propagation speed squared is always close 
to 1 (slightly larger than 1), which means that
the Laplacian instability of the tensor perturbation 
can be avoided.

%%%%%%%%%%%%%%%%%%%%%%%%%%%%%%%%%%%%%%
\begin{figure}
\begin{centering}
\includegraphics[width=3.3in,height=3.2in]{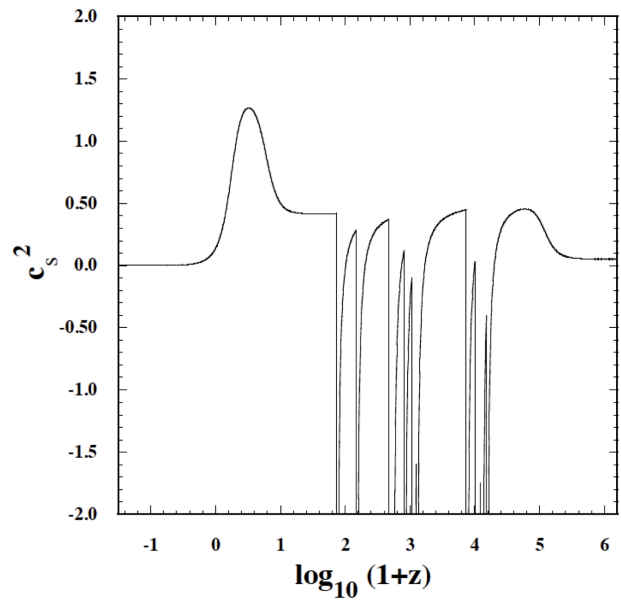} 
\par\end{centering}
\caption{
Evolution of $c_S^2$ versus $z$ for the model with $p=1$, 
$\epsilon_2=1$, $\epsilon_4=-1$, $d_4=1$, $d_5=-1$, $d_{\xi}=0$, 
and $\tilde{x}_{\rm dS}=-0.007$.
The initial conditions are chosen to be
$\tilde{x}=-4.5 \times 10^{-15}$, $\tilde{y}=-7.0 \times 10^{-5}$,
and $\Omega_{r}=0.996$ at $z=1.56 \times 10^{6}$.
In this case the scalar perturbation is subject to 
the Laplacian instability for many times by today.}
\centering{}\label{fig10} 
\end{figure}
%%%%%%%%%%%%%%%%%%%%%%%%%%%%%%%%%%%%%%

Let us next discuss the case with (ii), i.e. $p \tx<0$ and $d_5<0$
initially. In Fig.~\ref{fig10} we plot one example
for the evolution of $c_S^2$ with $p=1$, $d_5=-1$, and 
$\tx_{\rm dS}=-0.007$. In this case the density parameters 
as well as the effective 
equation of state evolve similarly as those in Fig.~\ref{fig6}.
However, even if the variable $\tx$ starts from negative values, 
$\tx$ crosses 0 for many times before reaching the dS solution 
with $\tx_{\rm dS}=-0.007$.
As we see in Fig.~\ref{fig10}, this leads to the violation of the 
condition $c_S^2>0$ by today. In addition 
the quantity $Q_S$ also becomes negative during some periods.
We have run our numerical code for many other cases in which 
the condition $p \tx_{\rm dS}<0$ is satisfied and found that 
in the case (ii) it is difficult to find a viable cosmological 
trajectory along which all of the no-ghost and stability conditions 
are satisfied.

In summary, we have shown that the cosmological solutions 
along which $p \tx>0$ and $d_5>0$ initially and $p \tx_{\rm dS}>0$
at the de Sitter attractor can evade the problems of 
the ghost and instability problems for $z \ge 0$.
In this case, although the scalar Laplacian instability 
is present at the de Sitter fixed point, the crossing 
of $c_S^2$ at 0 occurs at some time in future.
We have also run the numerical code for the initial 
conditions with $|\ty^2| \gtrsim |\tx|$ and found similar
properties of solutions to those discussed in this section.

\subsection{Initial conditions with $\ty^2 \ll |\tx|$}

Finally we shall study the case in which $|\ty^2| \ll |\tx|$ 
in the early cosmological epoch.
In this regime the term ${\cal L}_2$ is the dominant contribution 
to $\Omega_{\rm DE}$ relative to ${\cal L}_{3,4,5}$, i.e.
$\Omega_{\rm DE} \simeq -p \tx -\epsilon_2 d_2^2 p^2 \tx^2/24$.
The quantities $Q_S$ and $Q_T$ are approximately given by 
\begin{equation}
\frac{Q_S}{M_{\rm pl}^2} \simeq 
\frac{p^2 \tx^2 (6-\epsilon_2 d_2^2)}{(2+p \tx)^2}
\left( \frac{\phi}{M_{\rm pl}} \right)^p\,,\qquad
\frac{Q_T}{M_{\rm pl}^2} \simeq \frac{1}{2}
\left( \frac{\phi}{M_{\rm pl}} \right)^p\,,
\end{equation}
whereas both $c_S^2$ and $c_T^2$ are close to be 1.
The tensor ghost can be avoided for $(\phi/M_{\rm pl})^p>0$.
Under this condition the scalar ghost is absent for $\epsilon_2=-1$.
If $\epsilon_2=+1$, the absence of the scalar ghost 
requires that 
\begin{equation}
d_2^2 <6\,.
\label{d2con}
\end{equation}
For $d_4$ and $d_5$ of the order of unity we find from 
Eq.~(\ref{gened2}) that $\epsilon_2 (d_2 p \tx_{\rm dS})^2 \simeq 24$, 
where we used the condition (\ref{genecon1}).
Hence the dS solution exists only for $\epsilon_2=+1$, 
in which case $d_2^2 \simeq 24/(p \tx_{\rm dS})^2 \gg 1$.
This is incompatible with the condition (\ref{d2con}).

These results show that, if the solutions start from the 
regime $|\ty^2| \ll |\tx|$ with $\epsilon_2=+1$ (i.e. negative 
kinetic energy), the requirement for the avoidance of 
ghosts at the initial stage is not compatible with the existence 
of the dS solution at late times.

%%%%%%%%%%%%%%%%%%%%%
\section{Conclusions}
%%%%%%%%%%%%%%%%%%%%%

In this paper we have studied the cosmology of generalized Galileon
theories based on the Lagrangian (\ref{Lag}). For each Lagrangian
${\cal L}_{i}$ ($i=1,\cdots,5$) the scalar field $\phi$ is replaced
by general scalar functions $f_{i}(\phi)$. The covariant Galileon
theory satisfies the Galilean symmetry symmetry $\partial_{\mu}\phi\to\partial_{\mu}\phi+b_{\mu}$
in the Minkowski space-time. The extension to scalar functions $f_{i}(\phi)$
generally breaks this symmetry, but the equations of motion remain
at second-order. This is a welcome feature to avoid the propagation
of the extra ghost degree of freedom. We have also taken into account
two terms ${\cal L}_{6}=F(\phi)R$ and ${\cal L}_{7}=\xi(\phi){\cal G}$
that give rise to second-order equations and vanish in the Minkowski
space-time.

In the flat FLRW cosmological background we have derived the equations
of motion (\ref{eq:frd1})-(\ref{conser}) for the general Lagrangian
(\ref{Lag}). If we demand the existence of dS solutions, the functions
$F(\phi)$, $f_{i}(\phi)$, and $\xi(\phi)$ are restricted to be
either in the form (\ref{funchoice}) or (\ref{funchoice2}). The
former corresponds to the covariant Galileon theory with constant
$F$, respecting the Galilean symmetry in the Minkowski space-time.
The latter can be regarded as a kind of scalar-tensor theories in
which $F$ is field-dependent.

In the presence of two perfect fluids we have also derived conditions
for the avoidance of ghosts and Laplacian instabilities associated
with scalar and tensor perturbations. The no-ghost conditions (\ref{eq:constro1})
and (\ref{eq:constro2}) are automatically satisfied for the perfect
fluids of radiation and non-relativistic matter. Then the no-ghost
condition of the scalar mode is given by Eq.~(\ref{eq:genQs}), whereas
the ghost is absent for the tensor mode under the condition (\ref{eq:genQW}).
The stability conditions for scalar and tensor perturbations are given,
respectively, by Eqs.~(\ref{cS2}) and (\ref{eq:noinstGW}). We have
applied these results to two theories having dS solutions. For the
theory with constant $F$ the dS solutions are always classically
stable against homogeneous perturbations, whereas for the theory with
non-constant $F$ they are stable under the condition (\ref{dsstability}).

We have carried out detailed analysis for the cosmological dynamics
of the covariant Galileon theory with constant $F$. Introducing the
dimensionless variables $r_{1}$, $r_{2}$, and $\Omega_{r}$ together
with the constants $\alpha$ and $\beta$, it is possible to express
autonomous equations as well as physical quantities (both background
and perturbations) in terms of those variables in a convenient form.
In particular we showed the existence of an interesting tracker solution
$r_{1}=1$, along which the field velocity evolves as $\dot{\phi}\propto1/H$.
On this tracker all the non-linear field Lagrangians contribute to
the field energy density with the similar order, such that any of
these terms cannot be neglected. Moreover the cosmological dynamics
along $r_{1}=1$ does not depend on the parameters $\alpha$ and $\beta$,
see Eqs.~(\ref{eq:dR2}) and (\ref{eq:dOmr}). The solutions with
different initial conditions converge to a common trajectory, depending
on the epoch at which they reach the regime $r_{1}\simeq1$.

Along the tracker solution the dark energy equation of
state is given by Eq.~(\ref{wdetra}), which exhibits peculiar evolution:
$w_{{\rm DE}}=-7/3$ (radiation era), $w_{{\rm DE}}=-2$ (matter era),
and $w_{{\rm DE}}=-1$ (dS era). Since we have derived analytic formulas
for $w_{{\rm DE}}$ as well as $r_{2}$ and $\Omega_{r}$ in terms
of the scale factor $a$, this will be convenient to confront the
Galileon theory with supernovae observations.

Although the background dynamics on the tracker does not depend on
the parameters $\alpha$ and $\beta$, the conditions for the avoidance
of ghosts and Laplacian instabilities do. In Fig.~\ref{fig1} we
showed the viable parameter space in the $(\alpha,\beta)$ plane constrained
by the no-ghost and stability conditions along $r_{1}=1$. If the
solutions start from the regime $r_{1}\ll1$, we also require the
condition $\beta>0$ to avoid the scalar ghost. In this case the tensor
mode becomes slightly super-luminal. In the Minkowski space-time the
only solution to the field equation in Galileon theory with $d_{2}\neq0$
corresponds to $\dot{\phi}=0$, so that the super-luminal
propagation is absent.

We have also studied the cosmology based on the theories with 
non-constant $F(\phi)$ having de Sitter solutions at late times.
For the initial conditions with $\ty^2 \gg \tx$ we require that 
$d_5 p \tx >0$ in the early cosmological epoch. If $p \tx >0$, 
there are some viable cosmological trajectories 
along which the solutions
fulfill all the required conditions by today.
Such an example is given in Figs.~\ref{fig6}-\ref{fig9}, 
along which the quantity 
$p \tx$ remains to be positive.
In this case the scalar perturbation is subject to the Laplacian instability 
at the de Sitter fixed point in future ($c_s^2=-p \tx_{\rm dS}/27<0$).
If $p \tx <0$ initially, we find that the violations of the conditions
$c_S^2>0$ or $Q_S>0$ typically occur by today.
For the initial conditions with $\ty^2 \ll \tx$ the condition 
for the avoidance of ghosts in the early cosmological epoch 
is not compatible with the existence of the late-time de Sitter solutions.

The field-derivative couplings with the Ricci scalar $R$ and the
Einstein tensor $G_{\nu\rho}$ appearing in the terms ${\cal L}_{4}$
and ${\cal L}_{5}$ can lead to imprints on the dynamics of matter density
perturbations through the change of 
the effective gravitational coupling.
It will be of interest to study the evolution of perturbations in detail
in order to discriminate between the generalized Galileon model and
other dark energy models.

%==================================================================

%%%%%%%%%%%%%%%%%%%%%%%%%%%%%%%%%%%%%%

\begin{acknowledgments}
The work of A.\,D.\ and S.\,T.\ was supported by the Grant-in-Aid
for Scientific Research Fund of the JSPS Nos.~09314 and 30318802.
S.\,T. also thanks financial support for the Grant-in-Aid for Scientific
Research on Innovative Areas (No.~21111006). We thank Savvas Nesseris
and Jiro Soda for useful discussions. 
\end{acknowledgments}
%%%%%%%%%%%%%%%%%%%%%%%%%%%%%%%%%%%%%%

%==================================================================

\end{document}